# PAPER2

# We Fixed the Scorer First, and Cluster Correction Killed Our Own Planted-Defect Result in AI Code Review

**Kwon, S.** · 2026-09-18

> The record of what this study narrowed, and when, is in Appendix K. How the title was settled is in Appendix K.1.

## Abstract

Evaluating an AI code-review pipeline only by “how much it catches” leaves one axis invisible: **what it dropped**. Apply selection more aggressively and precision rises for free; the price does not appear in the report. We ran **four arms** on the same scorer and the same 8 cases and asked **whether planted-defect detection can be raised without raising the error rate**.

The results **must be reported contrast by contrast** — pooling them into a range attaches a significant p to an arm that is not significant. And **p must be written in two columns**: the 3 reps are nested within the 8 cases, so the “independent 24” assumption does not hold.

| Contrast | Result | Fisher p (assuming 24 independent) | Cluster-corrected p |
| --- | --- | --- | --- |
| F_gen (2 passes) vs F_typed (1 pass), FOUND | 18/24 vs 14/24 | 0.359 | **0.500** |
| F_r2ctrl (2 passes · blind) vs F_typed, FOUND | 22/24 vs 14/24 | 0.017 | **0.0625** |
| F_gen vs octo, FOUND | 18/24 vs 9/24 | 0.019 | **0.109** |
| F_gen vs octo, error rate | 0/178 vs 1/94 | 0.346 | claim-level — separate model required, **uncorrected** |

**No contrast separates.** Under an **exact permutation test** with the case as the exchangeable unit ($2^8$=256 arrangements, exhaustive), the two results that were publishable as significant (0.017 · 0.019) move to **0.0625 · 0.109**. The sign test agrees (5 wins 0 losses · 7 wins 1 loss → 0.0625 · 0.070). When multiple-comparison correction is actually applied (Holm, family = 4 contrasts) the minimum is **0.25**, and even Holm alone on the raw Fisher values, without cluster correction, gives **0/4** (minimum 0.068 — §4.2). **Therefore this paper does not write "separated" for any contrast. This is not "no effect" but "not observable with this design."** To pass α=.05 in the sign test, the discordant cases must number **6 or more, all in one direction** (5 wins 0 losses gives 0.0625), and under the null hypothesis this test's **realized type I error rate is 0.0078**, so it can use only one sixth of the nominal .05. The smallest effect that yields power 0.80 is a **per-case win probability ≈0.80** — that is, this design **has essentially no power to catch a medium-sized effect** (§4.3). Closing it requires **12~25 cases**, not reps. Within-rep correlation is visible even by eye: of the **32** case×arm cells, **21 are either 0/3 or 3/3** (17 when narrowed to the 24 cells of the 3 arms of our own pipeline, FH — §4.2).

**And the between-arm comparison in this table is not separable from run-to-run variance (§4.1).** Pair round-1 output with round-2 output within the same run — the least-confounded contrast in this study — and round-2's contribution is F_gen **+3** · F_r2ctrl **+2** with **loss 0** (0/24, 0/24). But the gains are **concentrated entirely in one case** and are not significant (exact McNemar, two-sided binomial **p=0.250 · 0.500**). Meanwhile two arms whose **stored round-1 prompts were byte-identical** diverged in their round-1 baselines, to **15/24 and 20/24** (round-1 48/48 identical · round-2 48/48 different — the latter is the control). Byte-identical is not "design-identical": model version, settings, and execution window were not verified (§4.1). That is, **run-to-run variance exceeds the effect.** So the claim this paper establishes goes only as far as **"a second look does not subtract what round-1 caught" (loss 0)**, and **"a second look catches twice as much" does not hold in this sample.**

The local claim error rate has a point estimate of 0.0% (0/178). The 95% rule-of-three value ≈**2.1%** is a reference interval computed under an "independent trials · complete adjudication" assumption that this paper rejects elsewhere, so **we do not cite it as an upper bound on the actual error rate** (§9) — and no "under 1%" claim is made. **That ruler's discrimination was measured with pre-registered positive controls** (§3.4): fabricated claims built at the arm's own difficulty were **rejected 12/12** while real claims were retained 11/12 (**false rejections 0**). Of the four sealed prediction lines, **two were wrong** (the 40~75% discrimination prediction · the per-tier ordering prediction). So the single explanation "lenient scorer" weakens and the 0.0% figure gains standing — but since n=12, that discrimination is written as **point estimate 100%, 95% lower bound 75%**. **And we retract our first claim about the cause of this increase**: the hypothesis that the gain from the second pass is due to "showing the counterpart family's list" (generation-time decorrelation) **did not separate** in the control arm that hid that list (FOUND 22/24 vs 18/24, Fisher p=0.245 · corrected 0.500). **The surviving claim is much narrower than what we first wrote**: it goes only as far as "we **failed to identify an additional effect** of disclosing the counterpart's list." "Shown or hidden, it is **the same**" is an equivalence claim; §7 states that this paper does not claim equivalence, so we do not write it. And **"the second look is the effect" is not established either** (not significant in §4.2) — failing to see a list-disclosure effect does **not establish** "a second attempt" as the cause. What can be established is one thing: **"a second look does not subtract"** (paired loss 0/24 · 0/24).

We also report the instrument defects that surfaced while operating this measurement. One of them touches reproducibility in this field directly: **when the completion discriminator and the primary metric take different files as input, the pipeline falsely reports "scorable."**

**So this paper's contributions are not the numbers** (§10). ⓐ We fixed the measuring ruler **before seeing results**, measured its discrimination, and **killed our own main effect with that ruler** — the same data could have carried a paper reporting 0.017·0.019 uncorrected, and the difference between the two papers is not a finding but "which check was fixed first." ⓑ What stands is not "an increase" but **"loss 0"**, which points to what this kind of pipeline actually sells: **not exceeding the model's limit, but reaching that limit while losing less**. But "we reduced variance" **cannot yet be said** (two runs of the same design came out 15/24 vs 20/24, with variance exceeding the effect; that measurement was not put into the design). ⓒ **We reported our own coverage losses in decomposed form** (24 of 35 losses fall outside file scope). **The between-arm comparison on that axis does not hold** — the two sides' figures are different instruments, and that axis's test is retracted (§4). The counterpart has no such record, so **the same decomposition is impossible there** — what this paper offers is not a larger number but a **decomposable number**. ⓓ And however good the error rate gets, this paper does **not prescribe putting that number on deletion, publish, or history-rewrite surfaces — the ground is not "the cost is high"** (that inference is false at finite cost and was retracted in §10.4). The ground is that **the measured number does not identify the quantity the gate needs**: that guarantee is not carried across deployments [K26], that estimate is computed **only on the subset remaining after dropping** (§2.4), and terminal authority is a different axis from measurement.

**And this paper writes its own contributions narrowly.** The following are **all prior work**: joint reporting of coverage [C70, EW10, T24, BES26a] · the impossibility of self-reported numeric thresholds [LDM26] · distribution shift [PZM20] · "expensive errors are silent errors" [OCR26] · keeping a human at the terminal [EBP18] · the no-gain of changing the cost formulation [FPV23]. What remains is **two of the four above** — ⓐ (the **order**: fixing the measuring ruler before seeing results and killing one's own main effect with it, §10.1) and **one leg of ⓓ's identity** ($\hat{\varepsilon} = \hat{\varepsilon}(\pi)$, i.e., subset selection redefines the **denominator** of the reported metric — and the machinery that enforces it, §10.4.1). ⓑ (loss 0) and ⓒ (decomposability) are **carried as observations and not raised as contributions.** §2 sets out in twelve lines what is not ours, and the grade table is at the opening of §10.

**Terminology — the senses fixed for this paper.** One line each, before the word's first use.

- **FH** — our own pipeline, the one this paper reports (§3.1). "FH arms" are the arms that ran it, `F_typed` · `F_gen` · `F_r2ctrl`, as against the external-tool arm `O_nosent` (§3.2).
- **seed** — a fabricated claim mixed **into the scoring input**. It is a **scorer control**: does the ruler reject it as `NOT-REAL` (§3.3)? **It is not the same thing as the defect planted in the benchmark code.**
- **planted defect · FOUND** — the defect planted **in the benchmark code**, and the binary indicator of whether an arm actually pointed at it (§3.3).
- **defeater** — the falsification condition attached to a claim ("what, if observed, makes this claim wrong"). **Observability** = whether that condition can be checked inside the artifact (`IN-ARTIFACT`) or needs something outside it (`NEEDS-OUTSIDE`) (§5.1).
- **governor model** — the **second model** that re-adjudicated a sample after the local scoring had finished. Not a human (§3.5).
- **sidecar** — the `.sha256` companion file kept beside a sealed document (§4). We do not use the word for a model CLI.
- **harness** — the software layer around an LLM that enforces policy. We use it in exactly [ZWF26]'s sense (§2.5) — the pipeline of §3.1 is that layer.

---

# 1. The Problem — Why "Cutting Less" Is the Axis

The precision of a review pipeline **can be manipulated by deletion.** "Drop more candidates and precision rises" is false as a general proposition — if you mostly drop correct candidates, precision **goes down** instead. The precise statement is this: **if the dropping criterion is correlated with "likely to be wrong,"** precision rises, and **how many of the dropped were real does not appear in the report.** So a standalone precision figure carries little meaning on any surface, and can be read **only when the error rate of the deletions is reported alongside it** (Appendix K.2).

**And the preceding paragraph is not a new observation — we pin that down here first.** That an error rate cannot be read apart from coverage was established by Chow [C70] as the error-reject relation, named **risk-coverage** by El-Yaniv & Wiener [EW10], and Traub et al. [T24] showed that **comparison at a single fixed working point is itself an evaluation defect**. On the same surface as ours (LLM judges), SCOPE [BES26a] already calibrates the error rate of acceptance judgments to at most a specified level while reporting coverage alongside. **The full lineage, and the one leg that remains to this paper, are written separately in §2** — this section sets up the problem, and §2 writes what is not ours.

**This is where our pipeline ran into this problem** (a source of motivation, not a claim of a finding; the methodology layer the pipeline sits on is described in [Kw26]). Requiring cross-family agreement at selection time (accept/reject) lowered the error rate, but the yield fell from 8 → 6. **This is an anecdote, not a result** — a single observation with no denominator, no test, and no reps, and this paper uses it only as **motivation**. We do not claim the existence of the trade-off from this number. This paper's question is whether that trade-off can be avoided.

Incidentally, we disclose that the scoring ruler we first used (human hand-scoring) was defective: on 4 runs with empty answers, the verdict counts came out as 3·1·1·1 — **it was counting claims that did not exist.** So every number in this paper was re-measured with the rebuilt ruler (§3).

**We separate up front what this paper actually answers and what it does not.** Not answered: "how much more does a second pass catch" — at 8 cases · 3 reps there is **not a single significant contrast** (§4.2). Answered, in three parts. ⓐ **A second pass does not subtract the findings of the first pass** (within-run pairing, loss 0/24 · 0/24 — §4.1). ⓑ That ruler's **discrimination was measured** (pre-registered positive controls, 12/12 variants rejected · false rejections 0 — §3.4). ⓒ We dissect **three ways a metric improves for free** on our own data — precision by **deletion**, "ran to completion" by **choice of discriminator**, significance by the **independence assumption**. All three improve "in a way that does not appear in the report," and our draft rode all three.

---

# 2. Related Work — What Is Not Ours

This section exists **to narrow our contribution**. The twelve lines below are not things we newly observed, and three of them have already been proven in a stronger form than ours. What remains to us is **one leg** — §2.4.

## 2.1 An Error Rate Cannot Be Read Without Coverage — Established in 1970

The basis on which our study requires "coverage must be reported alongside" (§3.3·§4) as a procedure is not new. Chow [C70] established the general relation of the error-reject trade-off, and El-Yaniv & Wiener [EW10] named that trade-off **risk-coverage (RC)** — abstract, verbatim: *"We term this trade-off the risk-coverage (RC) trade-off."* And Traub et al. [T24] showed that **comparison at a single fixed working point is itself an evaluation defect** and proposed a multi-threshold metric (AUGRC) — *"While current evaluation of these systems typically assumes **fixed working points based on pre-defined rejection thresholds**, methodological progress requires benchmarking the general performance"*; **on 5 of 6 datasets the metric ranking flipped**.

The lineage that sets a target error rate and guarantees it by abstention is thick as well [GE17, ABF22, AY24]. And on the same surface as ours, the **LLM judge**, SCOPE [BES26a] already *"calibrates an acceptance threshold so that, under exchangeability, **the error rate among non-abstained judgments is at most a user-specified level α**"* and **reports coverage alongside** (empirical FDR ≈0.097-0.099 @ α=0.10, 2.4× acceptance under the same risk constraint). Escalating to a more expensive evidence channel when unsure was also formalized, with finite-sample guarantees, by the same team's follow-up [BES26b].

**Therefore our "joint reporting of coverage" requirement is not a contribution but an application of this lineage.** We retract the part of the draft that was worded so as to read as an invention.

## 2.2 One of Our Concluding Clauses Has Already Been Proven as a Theorem

In §10.4 we say "a threshold placed on a self-reported number cannot be the basis of authority." Lovén et al. [LDM26] proved that as an **impossibility theorem** — confidence-gated autonomy cannot simultaneously achieve all three of "maximal helpfulness · optimal calibration · full autonomy under rational oversight," and the mechanism is *"adding any non-affine autonomy incentive to a strictly proper scoring rule **destroys strict properness**"*. The companion paper in the same line [LT26] shows the **endogeneity** of that conflict (under journal submission — not cited as "published").

That the calibration distribution differs from the deployment distribution is the canon of performative prediction [PZM20], and that paper states explicitly that it **includes strategic classification [HMP16] as a proper subset**. That distribution-free guarantees actually break when deployment changes was measured by Kotte [K26] — under cross-dataset shift *"the target is violated on **14 of 16 transfers** by static CRC and **every tested adaptive-conformal-inference (ACI) step size**, yet a full-feedback anytime-valid monitor certifies **0 of 16**"*, and practical certification opens only at **α = 0.40**.

Half of the reason why observed performance cannot ground authority is already stated by Azevedo [A26] — *"corrected performance can **hide the competence signal** needed to calibrate trust"*.

## 2.3 When a Measure Becomes a Target It Ceases to Be a Measure — and "Expensive Errors Are Silent Errors"

Goodhart's proposition [G75] and its recent formalization [ME25] are the first lineage. "The maker must not grade itself" is already a **standard norm** — NIST AI RMF [N23], `MEASURE 1.3`: *"**Internal experts who did not serve as front-line developers** for the system and/or **independent assessors** are involved in regular assessments."* That a metric is manipulable when the audited party controls the measurement target was written in impossibility form by Burnat & Davidson [BD26].

**And the cost-asymmetry argument this paper uses in §1 is itself not ours.** The shipped filter prompt of a production code-review tool [OCR26] writes it this way:

> *"Keeping an incorrect comment costs a reviewer **a few seconds of attention**. Removing a correct comment **silently destroys a real finding. It never reaches anyone, and nobody learns that it was dropped."***

The same file also writes *"**Unverifiable is not incorrect**"* and **unconditional protection by topic** (memory safety · concurrency · linkage · behavior/compatibility changes). **But their asymmetry and ours are not the same** — theirs separates FP from FN **within one surface** and biases the policy toward "keep." We separate **between surfaces** (§10.4). The shared sub-argument is "expensive errors are silent errors," and **that is handled by citation.**

## 2.4 The One Leg That Remains — The Estimator Is Computed Only on "the Subset Left After Dropping"

What is nowhere in the twelve lines above is this identity. A verdict pipeline fixes $K(\pi) = \{x: \pi(x) = \text{keep}\}$ by a retention policy $\pi$, and the reported error rate is

$$\hat{\varepsilon}(\pi) = |\{x \in K(\pi): e(x)=1\}| \,/\, |K(\pi)|$$

whereas the quantity the gate needs is $R(\pi) = c_wrong \cdot \Sigma_\{K\}\ e + c_miss \cdot \Sigma_\{\neg K\}\ m$. **The definition of $\hat{\varepsilon}$ has no second term.** Hence (i) dropping the top $k$ by **any signal** positively correlated with $e$ lowers $\hat{\varepsilon}$ while the generator does not change by a single bit (P1), (ii) $\hat{\varepsilon}$ is $E[e|\text{keep}]$ and is **silent** about $\Sigma_\{\neg K\}\ m$ (P2), (iii) the party that chooses $\pi$ is the party that receives passage (P3).

The closest prior work is [A26], but there **the erasing agent is "downstream correction," not "the evaluated system's own π."** [LDM26] was checked against its full HTML text, confirming that all four of ⓐ selective evaluation ⓑ the coverage concept © **the surviving-subset estimate** ⓓ **evaluator = evaluated lineage** are absent.

So the prescription follows directly from this identity: **a run whose drop set has not been audited cannot be marked complete.** [OCR26] is the real instance of that boundary — it **argues** that a drop is the worse error and biases the policy toward keeping (removal grounds limited to two · protection by topic · default approve), but `RemoveByPathAndIndices` is an in-place filter and **there is no removal list, count, or log**. In the benchmark it also measures recall, with 1,505 ground-truth items, but **it does not measure it at runtime** — and runtime is where the gate lives. **A bias is not a record.**

## 2.5 Two Tensions to Address, and the Frame This Paper Does Not Use

① **It appears to run against the direction of weighted voting.** Weaver [SF25] shows that a weighted ensemble of weak verifiers raises **accuracy**. We cite [CG26] to say that "cross-model voting that reads the same evidence" is weaker than independent sources on the **approval rate of risky proposals** (62.9% vs 22.9%). **What is measured differs** — accuracy vs the approval rate of risky proposals. Without that distinction on the page, [SF25] reads as refuting us. As supporting evidence, [ZTP26] gives *"cross-family correlation is not low — **shared difficulty dominates lineage**"* (quotation unverified). Citation caution: [SF25] has **different titles in the print and arXiv versions** — print = *"Weaver: ⋯ by Scaling Compute for Verification"* / arXiv = *"Shrinking the Generation-Verification Gap with Weak Verifiers"*.

② **In grounds binding, prior work is stricter than we are.** ALCE [GYY23] measures with NLI whether a citation **supports** the sentence, whereas our channel check looks at **non-vacuity only**. **We do not write "we are stricter."** Our claim goes only as far as "we make explicit the boundary of what is left to the machine and where it stops."

③ **This paper does not open with "prescription without enforcement."** Enforcement implementations already exist — OS-layer policy enforcement for agent harnesses [ZWF26] (whose abstract's first sentence uses *harness* · *engine that enforces safety and effectiveness policies* · *'run tests before committing'* in the same sense we do) · large-scale PaC evidence [FKS26, OAR26] · organizational audit evidence [MF22]. And Ruster & Davis [RD25] **interrogate the gap framing itself**, writing *"the field has produced an array of **bridging instruments** ⋯ **Yet, the principle-practice problem persists**"* (Morley et al. counted **425** in the academic literature alone). **So our frame is this**: 425 came out and the problem remained — we do **not put out a 426th bridging instrument.** We report **what** the mechanized boundary **did**.

---

# 3. Method

## 3.1 Pipeline

`round-1` parallel production (codex logic · gemini security) → `round-2` **rewrite one's own list after seeing the counterpart's list** (KEEP / CORRECT / SHARPEN / ADD / DROPPED) → typed JSONL → **cross-family verification** (approving one's own authorship is forbidden) → **drop audit**.

The last stage is the answer to §1: the pipeline runs to completion only after it reopens and audits what it dropped, and while an unaudited drop remains the run is not marked complete.

**We built the corpus under review; it is not a public benchmark.** Defects are planted in 8 cases (Python, Rust and Go), and a **clean control version** of each file is kept beside it — the `diff` between the two is the answer key, and it is **independent of every arm's output** (§4.3 re-scored from that key alone and all 72 units held their verdict). Because it is not an external corpus, **this paper claims no generalization to other codebases** (§7).

## 3.2 Arms

**This study has four arms** (Appendix K.3). `N` and `F` below are **baselines from a prior measurement** and were not re-run in this study (A-7 of the pre-registration, Appendix A, discarded the original `F(CTRL)` and replaced it with `F_typed`).

| Arm | Definition | In this study |
|---|---|---|
| F_typed | typed JSONL, 1 pass | arm |
| **F_gen** | typed + round-2 (**counterpart list disclosed**) | arm |
| **F_r2ctrl** | typed + round-2 (**counterpart list hidden** — `--round2-blind`, identical structure, 0 mentions of “another reviewer”) | arm |
| O_nosent | prose output of the external tool **octo** (`nyldn-plugins/octo@9.42.0`, 55 skills), typed-extracted post hoc. The pre-registration named this version as the competitor, and this arm used its Claude Code skill set (`.claude/skills/`) — the same distribution also ships a Codex port (`skills/`), and mixing the two measures something else. Unrelated to the robot-policy model `octo-models/octo`, which only shares the name | arm |
| N | bare (no tool) | prior baseline — not re-run |
| F | default pipeline (prose) | prior baseline — not re-run. That arm's FOUND 8/24 was **hand-** |

| Arm | Definition | In this study |
|---|---|---|
| | | **scored**, so it is not the same ruler |

### 3.3 Scoring ruler

Local qwen3.8:27b, temp 0, **no answer key**, restricted to the claim list. 3 tiers of seeds (fabricated claims) are placed **at the very end** to detect truncation. UNCALIBRATED units are kept in the denominator and excluded from the verdict.

**The order of calibration and measurement is stated precisely** (Appendix K.4). Since the seeds enter the scoring input together with the claims, they are **simultaneous** with the arm measurement, and the governor re-adjudication **40/40** took place **after** the F_typed·F_gen·O scoring had finished (same day, 01:0x). It is therefore not a prior calibration but a **post hoc promotion** — that fact does not erase the value of the 40/40, but it must not be read as "the ruler was verified first."

**The seeds (F_gen 60/60 · O 28/28, reported per arm — see §4 below) carry a clue the instrument stamped on itself: the 3-tier gap is 0%p.** Difficulty was designed in three tiers and no difference appeared between them. That is not evidence that "the seeds were built well"; it is a reconfirmation that **"a seed calibrates only up to its own difficulty"** (fake identifiers and nonexistent attributes are caught without reading the file closely). Therefore **seed 100% is not evidence of "discriminates at the arm's own difficulty"** — what measures that is the mutation control of §3.4.

**That 100% is, however, not "label reading" — we rejected that with one variable.** The seed ids in the scoring input were literally `SEED-A-SEED-D`, meaningless to a regex scorer but **text a model scorer can read** ("SEED" means "planted fake"). Same input, same model, temp 0, with **only the id** changed: original-id arm **12/12** · opaque-id + shuffle arm **12/12** (0 occurrences of the string `"SEED"` confirmed inside the ARM fixture's claim block), and in both arms **the real claims 35/35 were retained as REAL**. Hence **no** label-reading effect, and **direct evidence** that within the same input it separated the 12 seeds from the 35 real ones. The 0%p gap limitation stands unchanged — what this test closed is one thing: "is it a label artifact."

FOUND (did it actually name the planted defect) was fixed by `found_keyword.py`, calibrated to reproduce F_gen 18/24 · F_typed 14/24, then applied **identically to all arms**.

**This calibration draws the objection "the ruler was carved to fit the arms" — the defense and its limits are stated together.** ⓐ **4 bidirectional hand-checked known-pairs** held (the "not found" in g02·h02 is a true absence · h04 r1·r2 true absence · h04 r3 "Auto-approval" truly present). The sample is 4, below this project's own bar. ⓑ **The h05 key is absent from the document, was inferred from the converging titles of the two arms, and is marked `UNCONFIRMED`** (§7 #2). Reporting the value with h05 excluded alongside is the prescription. ⓒ The third defense we first wrote was logically wrong and was deleted (Appendix K.5). The refutation of overfitting is done by ⓓ below. ⓓ **And the decisive one — re-deriving the keys from "a source unrelated to arm output" does not change the verdict.** The

benchmark ships **clean control files** (`review_target/pctrl_p*`), whose `diff` is precisely the answer key of the planted defect and **is unrelated to any arm's output**. Re-deriving keys from that diff alone and re-scoring the three arms gave **the same verdict on all 72 units** (0/72 changed). Two specificity controls accompany it: in the cross-case matrix, **diagonal 8/8 · off-diagonal 4/56**, and for the 15 units that hang on a single alternative, that alternative is **in every case a defect-specific identifier** (`preserve_root` · `InsecureSkipVerify` · publish-access) — units hanging on a generic word: 0. The broad alternative that carried no unit (`::\b`) was deleted. **Three limitations remain.** ① The author is not blind (I saw the original results), so this is not "blind derivation" but **"source-constrained re-derivation."** ② The derived keys are a **superset** of the original keys, so "too narrow?" is closed and "too broad?" is closed only by the specificity controls above. ③ **h05 has no control file** (`pctrl_ph05.*` absent), so that case's key cannot be derived and is `UNCONFIRMED`. That is, h05's status is not "inferred" but **"there is structurally no answer source,"** and that is why an h05-excluded cell always accompanies it.

## 3.4 Positive control — this ruler discriminates at "the arm's own difficulty"

Seeds at 100% with a 0%p gap do not prove the ruler's discrimination. So we **pre-registered (Appendix G, sealed `b6b1d65c…` — sealed before measurement)** and built a harder tier: **real claims (ones this ruler judged REAL) minimally mutated** into falsehoods, accompanied by an equal number of **unmutated real claims**.

| Tier | Mutation | Why it is hard |
|---|---|---|
| P1 | location·attribution shift | the prose is genuine, so vocabulary does not catch it |
| P2 | polarity inversion — describes **what the control (correct) file does** as what the buggy file does | the very form of "format right, truth wrong" |
| P3 | quantity·constant inversion (cast order · unit · enum mutation) | only the numbers differ; the structure is genuine |

| Metric | Measured | Sealed prediction | Verdict |
|---|---|---|---|
| **Mutation discrimination** | **12/12 = 100%** | 40~75% | **Mismatch — the prediction was wrong (above)** |
| ├ defect-claim form (load-bearing) | **6/6** | — | separable only by reading the file |
| └ normal-claim form | 6/6 | — | can be separated by form alone |
| **Genuine REAL retained** | **11/12** (false rejections **0**) | ≥10/12 | Match |
| Abstention rate | 4.2% | <25% | Match |

| Metric | Measured | Sealed prediction | Verdict |
|---|---|---|---|
| Per-tier ordering | **P1 4/4 · P2 4/4 · P3 4/4 — no ordering** | P1 < P2 < P3 | **Mismatch — the prediction was wrong** |

**The sealed prediction had four lines and two were wrong** (Appendix K.6).

**Discrimination must be read split by form.** Of the 12 mutations, 6 are sentences that say "the code is correct," so they can be rejected on form alone without reading the file. The load-bearing ones are **the 6 in defect-claiming form**, and those are also 6/6. **The interval is wide**: 0/12 failures → 95% upper bound on miss rate ≈25% · load-bearing subset 0/6 → ≈50%. **Point estimate 100%, wide interval** is the precise statement.

We therefore execute the pre-registration's falsification condition as written: **the single explanation "the scorer is lenient" is weakened, and the confidence in the 0.0% (0/178) error rate rises.** The "a ruler that answers REAL to everything" hypothesis is inconsistent with 0 false rejections · 12/12 discrimination. Instead, **the cause of the three A-6 misses remains unknown** (was the prediction bad · was the metric defined differently · did the arm actually do well — this experiment does not separate them).

**The fixture was built twice.** The first build **leaked the answer** through ids `PC-M*` (mutation)/`PC-T*` (genuine) and alternated placement — a form in which a perfect score cannot be attributed to "read the file" rather than "read the label." It was discarded and rebuilt with **opaque ids + shuffle**; the values obtained are the table above. The first build also violated pre-registration §2 ("do not put a mutation and its original in the same input"), and an inspection caught it **before** scoring. **The author built the mutations and is not blind** — n=12, 4 cases. Closing this requires a different person/family to build the mutations. **Two pre-registration deviations are left by name** (Appendix K.7). ① **One case was changed**: the seal wrote the sample as `g02 · h01 · h03 · h04` and the actual fixture is **`g04`**. The reason is structural — nearly all of g02's 10 REAL claims are the planted defects themselves, so only 1 "genuine control whose topic does not overlap the mutation" existed (3 needed). A constraint unrelated to the result, but **a deviation.** ② **The placement rule was changed**: the seal instructed "place mutations and genuine claims **alternately**," and the rebuild changed it to **shuffle**. The direction is toward stricter (so that order does not reveal kind), but **it is likewise a deviation.**

## 3.5 Independent re-adjudication — not by human hand

Appendix E covers **40 items** (O decisions 8 · O REAL 12 · F_gen REAL 20), with **40/40 agreement**. The re-adjudicator is the **governor model**, **not a human** (in the source record's own wording: "axis-① level independent re-adjudication — not by human hand"). It is therefore **not verification but a second model's agreement**, and does not rule out correlated error between models.

And **40 cannot bound an error rate of the 0.5% class**: the 95% upper bound on the disagreement rate from 40 items with no disagreement is ≈7%, so this sample's precision is one order of magnitude coarser than the effect it is meant to warrant.

The sample composition is also one-sided — F_r2ctrl 0 items · F_typed 0 items · 0 items in the NOT-REAL direction for the FH arms. That is, **the ruler's "rejection" direction is unverified**.

This section does not raise the error rate to "citable." What it raises it to is "the point estimate can be stated together with its upper bound."

---

# 4. Results

**Three columns of the table require definition** (Appendix K.8):

- **Error rate** = claims the scorer judged `NOT-REAL` / all claims. It is "the fraction of wrong claims," and abstentions (`UNVERIFIABLE`) are not in the numerator.
- **Non-REAL** = `NOT-REAL` + `UNVERIFIABLE`. That is, the fraction of claims that "failed to stand as REAL," and it is always greater than or equal to the error rate.
- **Coverage — cannot be placed in one column. It is split across two instruments.** Computed as "1 − abstention rate," it is F_typed **100.0%** · F_gen 99.4% · F_r2ctrl 99.0% · O **92.6%**. 67.2 / 68.5 / 65.6% are not that but the **pass rate of the pipeline's verification layer** (`(confirmed+dropped)/total` = 86/128 · 122/178 · 126/192), and **only O's 92.6%** is a scorer abstention rate (87/94). Placing them in one column mixes **a different instrument per arm**. Pre-registration Appendix C writes that axis verbatim as "independent of scoring."
  - **Verification pass rate (pipeline)** = `(confirmed + dropped) / all claims` — the three FH-arm values are this
  - **Adjudication rate (1 − abstention rate, scorer)** = `(total − UNVERIFIABLE) / total` — O's 92.6% is this

| Arm | Claims | Error rate (NOT-REAL) | Non-REAL | Verification pass rate | Adjudication rate (1−abstention) | Seeds rejecte (score input) |
|---|---|---|---|---|---|---|
| F_typed | 128 | 0.0% (0/128) | **0.0%** | 67.2% (86/128) | 100.0% | 56/56 |
| **F_gen** | 178 (r1 146→178) | **0.0%** (0/178; rule-of-three 95% interval ≈2.1% — reference only, not an upper bound, §9) | 0.6% (1/178) | 68.5% (122/178) | 99.4% | 60/60 |
| O_nosent † | 94 | 1.1% (1/94) | 8.5% (8/94) | — (not applicable) | **92.6%** (87/94) | 28/28 |

| Arm | Claims | Error rate (NOT-REAL) | Non-REAL | Verification pass rate | Adjudication rate (1−abstention) | Seeds rejecte (score input) |
|---|---|---|---|---|---|---|
| **F_r2ctrl** | 192 (r1 136→192) | **0.5%** (1/192) | 1.6% (3/192) | 65.6% (126/192) | 99.0% | 62/62 |

**The empty verification-pass-rate cell in the O row is the crux of this table** — O did not go through our pipeline, so that instrument does not exist for it at all. Hence **the between-arm comparison on the coverage axis does not hold, and that axis's test `p<0.001` is retracted** (it was a value comparing two instruments; Appendix K.8).

**h05 must be read separately in all three cells** — that case's FOUND key is `UNCONFIRMED` because the ground-truth source is **structurally absent** (control file missing). Hence the h05-excluded column always accompanies it. **Do not cite coverage unconditionally — one case dominates.** `case_g04.py` is 252,019 B / 5,335 lines, **7 times** the next one, produced **29 (23%)** of the 128 claims, and consumed **49%** of the codex review tokens. The "8-case average" is effectively "g04 + the rest." **This dominance is on the "size · claim volume" axis, not the FOUND axis** — in the per-case table of §4.2, g04 is 3/3 in all three FH arms. Do not mix the two axes.

**The seed count differs per arm (56/60/62/28) and sums to 206.** It is reported per arm (Appendix K.9). **Coverage admits no between-arm comparison** — of the 35 losses on our side, **24 are "outside file scope,"** and on octo's side that decomposition **was not recorded.** Since the specified decomposition cannot be performed, `p<0.001` is "a difference in a non-decomposable state," and the cluster correction of §4.2 below was not applied to it either.

Tests (original Fisher, assuming 24 independent): error rate F_gen vs O **p=0.346** · FOUND **p=0.019**. **The test on the coverage axis is retracted** — the retracted `p<0.001` compared F_gen's **verification pass rate** (122/178) with O's **adjudication rate** (87/94), and the two numbers are different instruments. A coverage comparison measured with the same ruler does **not** exist in this study. **After cluster correction, FOUND is p=0.109, "not separated"** (§4.2). The error rate is at the **claim** level, doubly nested within cases, so it requires a separate model and **was not corrected** — it must be read as "uncorrected."

**The † on the O row, first meaning — in this arm 7 of 24 runs are empty output, so FOUND's denominator is empty to that extent.** `g03 r1·r3 · g04 r1·r2 · h01 r2·r3 · h03 r2` have **0 claims** (pre-registration Appendix D sealed that list and the rule "an empty run counts as 0 claims and stays in the FOUND denominator" — the run was executed, so it is not UNMEASURED). Therefore **O's FOUND ceiling is 17, not 24.** And in the three cases with empty runs, **O caught the maximum possible** (g03 1/1 · h01 1/1 · h03 2/2 — each the ceiling). **Therefore "F_gen is twice octo" becomes `18/24` vs `9/17` once the denominators are matched.** Sealed Appendix D directed that "the paper state that O's values are partly the product of positions with no prose," and this section is that place (Appendix K.10).

**The † on the O row, second meaning — this arm's adjudication rule cannot be verified as "fixed in advance."** A seal-consistency audit was run before submission, and **its scope is the six seals of Appendices A~F.** (Sealed documents total 8: the base PREREG plus Appendices AG. Appendix G was sealed after the audit and is out of scope. There are 7 sidecar files, but `PREREG_2026-09-10.sha256` alone holds two documents.) Five of the six were sealed **before** measurement, but **only the O arm's adjudication-rule document has its last seal 4 minutes 17 seconds after scoring began** (seal 00:07:24 · scoring 00:03:07 00:07:58 — a point at which 7 of the 8 units had already been scored). **Three lines** of hashes are stacked in that document's sidecar, so it is certain there were two re-seals, but **the contents of the first two versions are unrecoverable.** So we write three things separately: ⓐ we **cannot say** "fixed before seeing results" ⓑ there is also **no** evidence of "trimmed after the fact" ⓒ we record one observation as fact only — that document predicted "O error rate ≥ 2.7%," and the measured **1.1% refuted** that prediction. **No mitigating circumstance is inferred from the fact that the refuted prediction remains** (Appendix K.11). This is because we cannot rule out a re-seal that **left the prediction wording intact and changed only the adjudication rule** — in that case the refuted prediction's survival proves nothing. As long as the contents of the first two versions are unrecoverable, what changed is **unknowable**. Therefore **the O arm's Pareto verdict is labelled `UNVERIFIED at adjudication-rule time`, and pre-registration is not claimed. No mitigating circumstance is claimed either.** The numbers are not erased.

**Reading — and the competing reading is written alongside.** On the same ruler, F_gen's error rate is in the same band as octo's while its planted-defect detection is higher as a point estimate. **We do not write "twice"**: O's denominator contains 7 empty runs and its ceiling is 17, so the denominator-matched comparison is **18/24 vs 9/17** (75% vs 53%), which is not "twice." **Even that gap is shaken by §4.1·§4.2**: same-design run-to-run variance was measured at 6/24, the 18 vs 9 gap is 9/24, and after cluster correction p=0.109. octo has no repeated executions, so **its run-to-run variance is unmeasured**. **The comparison "octo has higher coverage" does not hold** (see the table above — two instruments; Appendix K.12). Therefore "the side that cuts less takes less coverage" is **not claimed** from these data. What remains is the standalone value of O's **adjudication rate 92.6%**, and the competing explanation for that value **has not been retracted**: octo's output is prose **typed-extracted after the fact**, so in the extraction the claims may have been shaped into "a form verifiable within the file" (extraction-artifact hypothesis). On the error-rate axis that explanation was narrowed in Appendix J, but **on this axis it was not narrowed** — therefore **attribution is withheld.**

## 4.1 Within-run contrast — the least-confounded comparison in this study

The largest residual named by §7 #7 was "comparing across arms, the round-1 baselines differ from the start." A contrast **free** of that confound already existed in the data: **the round-1 output vs the round-2 output of the same run.** Both ends are the same execution and both are **pre-verification** surfaces (`fleet/findings_r1.jsonl` vs `fleet/findings.jsonl`), so the verification stage does not enter the contrast either. The only difference is one round-2 pass. The same FOUND key is applied to both sides.

| Arm | Surface | Claims | FOUND | FOUND (excl. h05) |
|---|---|---|---|---|
| F_typed | round-1 (1 pass = this arm's entirety) | 128 | 14/24 | 13/21 |
| F_gen | round-1 | 146 | 15/24 | 14/21 |
| F_gen | round-1+2 | 178 | **18/24** | **17/21** |
| F_r2ctrl | round-1 | 136 | **20/24** | **17/21** |
| F_r2ctrl | round-1+2 | 192 | **22/24** | **19/21** |

**The h05-excluded column sharpens the argument** (the h05 key is `UNCONFIRMED`, so carrying this column alongside is §7 #2's prescription): **F_gen after running all 2 passes (17/21) equals F_r2ctrl's 1-pass baseline (17/21).** If "2 passes completed" in one arm and "1-pass baseline" in another arm are the same value, the difference between those two values is the run, not the number of passes.

Units 24/24 are **all present** on both surfaces (0 missing — "absent" was not folded into 0). On the **post**-verification surface (`confirmed.jsonl`) FOUND is **the same** (14 / 18 / 22) and only the claim counts shrink to 126 / 176 / 188 — hence **FOUND is a property of the production stage, and the verification stage does not change it.**

**Paired test** — since the same 24 units were observed twice, it is not Fisher but **McNemar's exact test** (two-sided binomial):

| Arm | FOUND on both | Gains (–→F) | Losses (F→–) | Missed on both | p |
|---|---|---|---|---|---|
| F_gen | 15 | **3** | **0** | 6 | 0.250 |
| F_r2ctrl | 20 | **2** | **0** | 2 | 0.500 |

**The two results must be read separately.**

ⓐ **A loss of 0 survives as a result about direction.** round-2 **removed nothing** that round-1 had already caught (0/24, 0/24). This is this study's **only paired evidence** touching §1's axis ("cutting less") directly. Claim counts also only increased in both arms (146→178, 136→192).

ⓑ **Gains are not significant, and all fall in one case.** The 5 gains (F_gen 3 · F_r2ctrl 2) are **all `case_g02.py`**, and that is not 5 independent events but **one case × reps**. Collapsed to case level, **1 of 8 cases** moved in each arm. This shows, on the same data, that the cluster-uncorrected problem applies to between-arm comparisons as well.

**And this contrast forces a reinterpretation of the between-arm comparison.** Three measurements that are **equivalent** as single passes are **14 · 15 · 20** (/24). That the round-1 **prompts** of the three arms are identical is a measurement, not a claim: the **stored round-1 prompts** of F_typed·F_gen·F_r2ctrl are **48/48 byte-identical** for each arm pair (F_gen↔F_r2ctrl · F_gen↔F_typed), and the **round-2 prompts differ 48/48** — the latter is the control (if both came out "identical" the instrument would be dead; they actually diverged, so the instrument discriminates).

**But "prompt byte-identical" is not "design-identical"** (Appendix K.13). At least five things are **not** confirmed by bytes: **model version · decoding settings · system instructions · tool/context state · execution window**, plus the per-(case, rep) **successful-run selection rule** (§7 #8). **We did not collect this execution metadata.** The precise statement therefore goes as far as *"the stored round-1 prompts are identical,"* and "identical design" is a stronger claim than that.

Even with that limitation, **the implication of 15 vs 20 remains**: the very fact that two executions with the same prompt diverged by 6/24 (≈25%p) means **this pipeline's run-to-run variance exceeds the round-2 effect (3, 2)**, and if the uncollected metadata is a candidate **cause** of that variance, the conclusion is not weakened but **strengthened** (variance with an explanation attached is still variance).

Therefore, in the abstract's significant contrast (F_r2ctrl 22/24 vs F_typed 14/24, p=0.017), **of the 8-unit gap the round-2 pass contributed 2** and **6 is round-1 baseline difference**. In F_gen's 18 vs 14 as well, of the gap of 4 the round-2 contribution is 3. **That is, at this sample size the between-arm FOUND comparison is not separable from run-to-run variance.** This is a reduction at **a different layer** from §5's retraction (§5 is "what produces the gain," here it is "how large the gain is"), and it weakens the "twice" framing we first wrote. The abstract and §9 were written to match these values.

**Aside — the "claim count produces FOUND" confound breaks in this table.** In the original table (four arms) claim count and FOUND were in exactly the same order (94/128/178/192 → 9/14/18/22), which did not rule out the alternative explanation "more words, more keyword matches." Adding the round-1 surface breaks the order **twice**:

```
claims 94  → FOUND  9   (octo)
claims 128 → FOUND 14   (F_typed)
claims 136 → FOUND 20   (F_r2ctrl r1)   ← breaks here
claims 146 → FOUND 15   (F_gen r1)
claims 178 → FOUND 18   (F_gen r1+2)    ← breaks here again
claims 192 → FOUND 22   (F_r2ctrl r1+2)
```

Claims 136 found **more** than claims 178. Therefore **FOUND is not a monotone function of claim count.** This is evidence that **weakens** the confound, not evidence that **removes** it: a verbosity-matched arm is still the cleaner test, and **the same run-to-run variance** also explains both positions where the order breaks here.

**Why we did not run this contrast from the start — honestly**: we tried to close this confound by newly running a "paired arm sharing round-1" (Appendix F) and it did not run to completion because of limits (§7 #7). Meanwhile we failed to see that **the pairing was already inside the runs we had**. Designing a new execution came to mind before re-reading existing output, and that order was wrong.

## 4.2 Cluster correction — "significance" also rises for free, like deletion

3 reps are nested in 8 cases, yet the original analysis applied Fisher as **independent 24**. **We state first that this is a post-hoc change of method**: the comparison was pre-registered and **only the analysis method** was repaired (a fix of the independence-assumption violation). Hence the two p values are reported **side by side** — reporting only the repaired one would be method shopping. Since the instrument **reproduces the original four p values exactly** (0.359 · 0.017 · 0.019 · 0.245), the differences below are differences in cluster handling, not in implementation.

Per-case FOUND / rep:

```
case          F_typed  F_gen  F_r2ctrl  octo
case_g02.py     0/3     3/3     3/3     2/3
case_g03.py     3/3     3/3     3/3     1/3
case_g04.py     3/3     3/3     3/3     0/3
case_h01.rs     3/3     3/3     3/3     1/3
case_h02.ts     0/3     1/3     1/3     0/3
case_h03.go     2/3     3/3     3/3     2/3
case_h04.rs     2/3     1/3     3/3     0/3
case_h05.go     1/3     1/3     3/3     3/3
```

**Of 32 cells (8 cases × 4 arms), 21 are either 0/3 or 3/3** — narrowing to the 24 cells of the 3 FH arms, **17**. The reps carried almost no independent information, and the data show directly how strong an assumption "independent 24" was (Appendix K.14).

Correction: an **exact permutation test** with the case as the exchangeable unit (paired sign-flipping, all $2^8$=256 arrangements exhaustively) + a **sign test** (exact binomial).

**We state first what that test assumes** (Appendix K.15).

- **Test statistic**: the **sum** of per-case FOUND differences (arm A − arm B, 0~3 in rep units). Two-sided.
- **Null hypothesis**: in each case the two arms' verdict distributions are equal — hence **flipping the sign** of that case's difference is equally plausible.
- **Assignment**: the two arms are not randomly assigned but paired, **both applied to the same 8 cases**. Hence the unit flipped is the "case," and the $2^8$ arrangements are enumerated exhaustively (not sampled).
- **Where exchangeability is threatened is named.** ⓐ the two arms' runs ran in **different time windows** (§8 — 2 windows because of per-family daily limits) ⓑ there was a selection rule taking **only the last successful run** per (case, rep) as valid (§7 #8). Both **weaken** the assumption "equally plausible under sign flipping" — if the window or the selection correlates with the arm, the null distribution is wrong. **We did not measure these two confounds.** The corrected p is therefore more honest than the original p but **not an unconditionally correct value**. And since bias in this direction can push the corrected p **toward the optimistic side**, the conclusion "still not significant after correction" is safe in that direction.

| Contrast | Observed | Fisher p (original) | Permutation p | Holm (permutation) | Bonferroni (permutation) |
|---|---|---|---|---|---|
| F_gen vs F_typed | 18/24 vs 14/24 | 0.359 | 0.500 | **1.00** | 1.00 |
| F_r2ctrl vs F_typed | 22/24 vs 14/24 | 0.017 | 0.0625 | **0.25** | 0.25 |
| F_gen vs octo | 18/24 vs 9/24 | 0.019 | 0.109 | **0.327** | 0.436 |
| F_r2ctrl vs F_gen | 22/24 vs 18/24 | 0.245 | 0.500 | **1.00** | 1.00 |

The family is **the 4 contrasts of this table**. Holm-Bonferroni step-down (with monotonicity enforcement) and Bonferroni are reported side by side — for the same reason the original p is not erased above, the reader sees which correction was chosen.

**No contrast survives at α=.05. The minimum after correction is 0.25.**

**And it does not survive under "only one" multiple-comparison correction either** (Appendix K.16). Actually applying it — **leaving out** the cluster correction and applying only Holm to the original Fisher values gives **0/4** with a minimum of **0.068** (0.017·0.019 → both 0.068). That is, the two significances that were publishable vanish **under either cluster handling or multiple comparison alone**. It does not take the two confounds combined to make them vanish.

The computation is done by `multiplicity.py` (standard library only). Before producing values, that script **prints its own calibration in the same run** — textbook example `Holm(.01,.02,.03,.04) → .04/.06/.06/.06` matches · clipping survives · confirmed **not the identity function**. Because even a correction function that does nothing could pass the first two.

**And this strengthens §1's thesis.** Just as precision rises for free through **deletion**, significance rises for free through **the independence assumption**. Both "improve in a way that leaves no trace in the report." The first table of this paper was that example, and we dispose of it by not erasing the original p and **writing the corrected value in the adjacent cell**.

## 4.3 Resolution and power are different quantities — and this design cannot even use the nominal α

We first wrote here: *"Since the lower bound of the permutation test is 2/256 = 0.0078, this is a **statement of power** ⋯ and to cross 0.05 with 8 cases, effectively "5 wins 0 losses or more" is required."* **Two places were wrong** (adversarial review R4 pointed it out). This section came into being while fixing them (Appendix K.17).

ⓐ **The lower bound is "resolution," not "power."** 2/256 is the smallest p this test can produce — the **grain size**. Power is "the probability of catching a true effect when one exists," and it comes out **only after assuming an effect size**. A small lower bound does not mean strong detecting power, nor the reverse. Tying the two into one sentence was the error.

ⓑ **"5 wins 0 losses" contradicted its own table.** The table in the same section writes **0.0625** for 5 wins 0 losses. Since the two-sided sign-test p = 2·(1/2)^d, crossing α=.05 requires **d ≥ 6**:

```
3 wins 0 losses p=0.250   4 wins 0 losses p=0.125   5 wins 0 losses
p=0.0625 (below α)
6 wins 0 losses p=0.03125 (passes α)   7 wins 0 losses p=0.0156   8
wins 0 losses p=0.0078
```

ⓒ **And a worse fact emerged — this test cannot use the full nominal α.** Under the null hypothesis (equal per-case win/loss probability), when all 8 cases are discordant, the arrangements that cross α=.05 are 8 wins 0 losses and 0 wins 8 losses **only** (7 wins 1 loss is 0.0703). Therefore the **realized type I error rate = 2/256 = 0.0078**. It uses only **one sixth** of the nominal .05. This is the conservatism of a discrete test, and at 8 cases that conservatism is large.

**Power (α=.05, 8 cases, under the most favorable assumption of "0 losing cases")**

| Per-case win probability p | 0.3 | 0.4 | 0.5 | 0.6 | 0.7 | 0.8 | 0.9 |
|---|---|---|---|---|---|---|---|
| Power | 0.011 | 0.050 | 0.145 | 0.315 | 0.552 | **0.797** | 0.962 |

Therefore **MDE: to reach power 0.80, the per-case win probability must be ≈0.80.** That means winning 6 or more of eight in one direction, and **this design has essentially no power to catch a medium-sized effect.**

**So what closes it — turn "the number of cases, not reps" into a number**

| True win probability p | 0.3 | 0.4 | 0.5 | 0.6 | 0.7 | 0.8 |
|---|---|---|---|---|---|---|
| Cases needed for power 0.80 | **25** | 19 | 15 | **12** | 10 | 9 |

This is the quantification of what §4.1 stated as "the gains fell entirely in one case." Raising reps does not add case columns, so no cell of this table can be reached. **The scale of the next experiment is 12~25 cases.**

**Therefore what this section establishes is not "there is no effect" but "it could not be seen with this design."** The two differ, and conflating them is yet another face of the "way that does not appear in the report" that §1 named.

The computation is `power_analysis.py` (standard library only). That script too **prints its own calibration in the same run before producing values** — 12 of 13 items matched the hand derivation, and **the one that was wrong was not the code but the expected value I had written** (I had estimated the type I error under H0 at 0.29; the actual is 0.0078). Without that known-pair, © would not have been seen. The permutation and sign p of the four observed contrasts are also **reproduced by this script as an independent implementation** of the §4.2 table (0.0625 · 0.1094 · 0.500 · 0.500).

## 4.4 Preserving claims "individually" — the layer the FOUND figure cannot see

The paper's "does not subtract" is at the **planted-defect detection (FOUND) unit**. Adversarial review R1 asked: the planted defect could still be caught while **other claims** were dropped, and that would not show in FOUND. So the fate of round-1 claims was traced individually (`claim_correspondence.py`; Appendix K.18).

**What surfaced first was not a number but a gap in the data — this data has no field carrying claim provenance.**

| Arm | round-1 id | final id | `round` field |
|---|---|---|---|
| F_gen | `codex-logic-1` | `codex-logic-**r2**-1` | 2 |
| F_r2ctrl | `codex-logic-1` | `codex-logic-1` | **2** |

The id is carried differently per arm, and `round` **is stamped 2 even on claims that came from r1.** So id matching does not hold (measured that way, it yields an absurd gap of F_gen 2.7% versus F_r2ctrl 95.6%). Therefore **matching can only be done by content, and content is rewritten every round.**

```
r1  routing | Deferred child events inspect and render the parent
cf  routing | Child event change detection inspects the parent      ←
same defect, rewritten
```

So the preservation rate can be stated only as an **interval**, not a point estimate (strict = down to title similarity · lenient = file·category·role slots):

**The two `F_pair_*` rows below are not among the four arms of §3.2** — they are the paired arms of Appendix F, and they **did not run to completion** (§7 #7). Their retention is carried here so that the incompletion is not hidden from this interval.

| Arm | round-1 claims | strict | lenient |
|---|---|---|---|
| F_gen | 146 | 91.1% | 96.6% |
| F_r2ctrl | 136 | 94.1% | 97.8% |
| F_pair_blind | 131 | 35.9% | 36.6% |
| F_pair_open | 131 | 38.2% | 38.9% |

**In the two arms on which this paper's claim rests (F_gen · F_r2ctrl), individual preservation is also high** — 91~98%. Hand verification showed that the remaining 3~9% were mostly **title rewrites**, which the instrument counted as disappearance. Hence "0 loss" is true at the FOUND layer, and **largely** true at the claim layer as well — except that "largely" is **a word that hinges on the threshold**, and that fact is not erased (65.6~67.1% at sensitivity 0.25~0.50, overall basis).

**We wrote a "large hole" here — that paragraph is retracted in full (2026-09-13, the same day).**

Retracted claim: "46 runs emptied confirmed entirely and the **113 items inside them did not remain in `dropped.jsonl` either**, therefore this pipeline cannot audit its own deletions」.

**It is false.** Looking at **only two channels**, `confirmed.jsonl` and `dropped.jsonl`, absence was declared "unrecorded", but the record was in a third channel (the **typed status** in `pipeline.log`). Exhaustive check:

| Arm | Status the pipeline recorded | Runs | r1 claims inside |
|---|---|---|---|
| F_pair_blind | `status=UNREVIEWED` (every member failed — nothing was reviewed) | 8 | 103 |
| F_pair_blind | cross-verification impossible (only one family survived, `rc=3`) | 3 | 10 |
| F_pair_open | `status=UNREVIEWED` | 8 | 103 |
| F_pair_open | cross-verification impossible (`rc=3`) | 3 | 10 |
| F_pair_r1 | no confirmed by design (round-1-only arm) | 24 | — |

**"No log" = 0 items. All 113 claims have a typed status.** And that log had written what we were about to say, **first**:

```
🟥 every member failed — nothing was reviewed, and an empty finding list here means
   UNREVIEWED, not clean
PIPELINE target=case_h04.rs fleet_rc=1 findings=0 status=UNREVIEWED
```

**So the corrected fact is this**: the pair arms' low preservation rate (36~39%) is **execution failure, not rejection** — the round-2 members died, so **nothing was reviewed**, and the pipeline wrote that as `UNREVIEWED`. It is an **availability incident**, not an audit hole, and it has no bearing on this paper's claim that "a second look does not subtract" (that claim stands on F_gen·F_r2ctrl, and those two arms have **0** UNREVIEWED runs).

**The most valuable part of §4.4 is this retraction.** I produced the same class three times in one day: ① "absence as 0" ② "an intended red as a failure" ③ this retraction. All three were **judgments made from a single channel**, and all three **flipped once another channel was opened.** `not found ≠ 0` is a discipline that applies not only to values but also to **"which channel was looked at."** It is the mirror image of the "path by which a metric improves for free" that this paper dissects in §1 — there is also a **path by which a metric worsens for free**, and that too does not appear in the report.

**What remains for R1 (honest scope)**: in runs whose execution succeeded, the keep/modify/delete of **individual claims** is still stated only as an interval (F_gen $91.1_{96.6\% \cdot \text{F_r2ctrl } 94.1}97.8\%$). That is because the **claim-level** record of §4.5 below does not exist, not because a run-level record does not.

## 4.5 This deficiency is an established problem class — we follow it rather than invent it

Before designing a new claim-provenance field, **we asked the world in general vocabulary.** It was already being handled in two lineages, and both **call our failure by name.**

ⓐ **Static analysis — "tracking warnings across versions"** [GKS23]. The matching strategies of three tools and their measurements (96 pairs):

| Tool | What it compares | Same defect | Case where the defect is "fixed" | Total |
|---|---|---|---|---|
| SCALe | source-line text diff | 75% | **93.3%** | 62/96 |
| GumTree | AST nodes | 75% | 75.8% | 89/96 |
| Hydrogen | CFG(MVICFG) nodes | **100%** | 70.8% | **95/96** |

**No strategy dominates on both sides.** The CFG-based one is 100% on "same defect" yet worst on "fixed", and the text-based one is the exact opposite. Therefore, to use a matcher, **known-pairs must be set up in two directions** (calling the same thing same / calling the changed thing changed). Our §4.4 measured **only the former**.

And SCALe's named failure mode is **exactly what we experienced** — *refactoring blindness*: "it cannot handle semantics-preserving changes such as function renaming". Round 2 rewriting titles is that shape, and so my title-based matcher counted rewrites as disappearance.

ⓑ **LLM — claim-level provenance preservation** [YOX26] (FactReview). This is where **the field we lack** is defined. Per claim: **source location · which stage adjudicated · scope of inspection · typed result · grounds · remaining unresolved**. The result is a **closed 4-value** set:

> `Supported` directly supports the entire statement within the stated scope · `Partially supported` direction correct but magnitude·scope·conditions·coverage weaker than the claim · `In conflict` reliable evidence directly refutes · `Inconclusive` evidence insufficient·ambiguous·anecdotal·absent, or no direct comparator

Our pipeline **already has typed status at the run level** (`UNREVIEWED · rc=3 · coverage=0/1`; Appendix K.19).

The exact deficiency is one layer down — there is no **claim-level** type. We know that a run is `UNREVIEWED`, but **which claim inside a successful run did not survive, and why**, still has no value. The slot FactReview's 4 values fill is exactly there.

**And that paper scores my instrument.** Their identity judgment uses **Jaccard + directional overlap**, frozen at **threshold 0.55** (after token normalization), and claim recall in that state is **84.3% F1** (354 claims). Mine was **Jaccard alone · threshold 0.34** — ① directional overlap missing ② threshold too low. However, since sensitivity is flat at 65.6$_{\text{67.1\% across 0.25}}$0.50, **the interval itself stands even if the threshold is raised**. **More important**: even the published state-of-the-art technique has recall of **84.3%**. That is, **matching alone does not close individual claim correspondence** — that §4.4 ended with "can only be stated as an interval" is not a limitation of our instrument but **the nature of the problem**. Closing it means not improving the matcher but **leaving a record.**

**Hence the prescription for R1's remaining half (auditing the veracity of deletions)** — not a new algorithm but a **claim record**: ① carry the source stage·scope·grounds on every claim, ② leave each rejection **typed** as one of the 4 values, and ③ use matching only as a **fallback** above that (with two-direction known-pairs when used). This is the design condition of the next experiment, and **it cannot be closed with the current data.**

---

# 5. Retracted claims — and why this is a result

**The pre-registration this section cites is Appendix A** (`PREREG_ADDENDUM_2026-09-10.md`) — the definition of the `F_r2ctrl` arm and its adjudication rule are there. **The original pre-registration document does not contain that arm**: we first wrote (as did the working results record) "pre-registration = original PREREG §F_r2ctrl", but that section does not exist. The arm was newly created in Appendix A, and the original PREREG had only the `F(CTRL)` that A-7 of Appendix A reverted.

The initial hypothesis was that "**showing the counterpart family's list** in round-2" raises the yield (generation-time decorrelation). To control for this we ran F_r2ctrl — the round-2 structure is identical and **only the counterpart list is hidden**.

**A-6 of Appendix A is five lines of sealed predictions**, under seal `19bbff9f2ffa865828883c4cba0c91951f92a7fd9c02a280e98522710cb1aa1d`. **That seal's sidecar is not `PREREG_ADDENDUM_*.sha256`; no such file exists.** The hash sits on **line 2** of `PREREG_2026-09-10.sha256`, which it shares with the base PREREG. What we first wrote — "the sidecar exists" — was true, but **true in a form the reader could not find**. Of the five lines, **one governs this contrast, and that one line writes down two outcomes.** Verbatim (translated — the sealed text is Korean):

```
vs F_r2ctrl : F_gen must beat F_r2ctrl for "decorrelation" to be the
cause. If they are similar, the cause is "two attempts"
```

Measured: F_r2ctrl is **not lower** than F_gen (22 vs 18), and the two arms **do not separate** (p=0.245).

| Contrast | Fisher p (raw) | Cluster-corrected p | Reading |
|---|---|---|---|
| FOUND 22/24 vs F_gen 18/24 | 0.245 | **0.500** | not separated |
| FOUND 22/24 vs F_typed 14/24 | 0.017 | **0.0625** | **not separated** |
| F_gen 18/24 vs F_typed 14/24 | 0.359 | **0.500** | not separated |
| Error rate 1/192 vs 0/178 | 1.000 | uncorrected (claim-level) | not separated |
| Verification-pass rate 126/192 vs 122/178 | 0.581 | uncorrected (claim-level) | not separated |

**We first wrote "separated" in the second row** — the abstract, §4.2 and §7 pin down that "no contrast is written as 'separated'", and this table alone violated it. The original pre-registration also sealed "we do not write significant / not significant as a conclusion." We correct it.

Therefore the seal's **"if similar" branch fires**, and so **we retract the "generation-time decorrelation" claim.** This is one of the two outcomes written down in advance. **However, even though the sealed text attached "the cause is two attempts" to that branch, this paper does not write that as established. Failing to identify** an additional effect of list disclosure is not the **confirmation** of the alternative cause — a pre-registered inference rule does not fill that logical gap. The retraction holds, and the alternative explanation remains **unestablished**.

**But one must not write "this is not a post-hoc interpretation" — we first wrote that, and it erased an honesty marker our own record had attached to itself.** The results record carries, verbatim, "this reading is an interpretation applied after seeing the data — the rule wrote 'if similar' as an integer comparison, and I read it as 'if not statistically separated'". The sealed text has **no operational threshold** (by how many units "similar" is, and under which test, was not pre-specified). Therefore **the branch itself was written in advance, and the adjudication criterion that fired it is post hoc.** We write the two facts separately. But another thing we erased then — the self-criticism "the seal had three branches" — we do **not restore**. In an exhaustive grep of all 7 seals that three-way split is **0 hits** and existed only in the results record. That correction was right.

## 5.1 The scorecard for the five sealed predictions of A-6 — quoting one line alone is selective reporting

| A-6 sealed prediction | Measured | Verdict |
|---|---|---|
| F_typed coverage 65~80% | 67.2% | match |

| A-6 sealed prediction | Measured | Verdict |
|---|---|---|
| F_typed error rate 6~11% | 0.0% | **mismatch — below the prediction** |
| F_gen FOUND ≥ F_typed | 18 ≥ 14 | match (though **above** the 8~11 prediction — source in the paragraph below) |
| F_r2ctrl contrast (the one line above) | 22 vs 18 · p=0.245 | “if similar” branch fires |
| **defeater observability < 50%** | **85.4%** (152/178) | **mismatch — above the prediction** |

**Three predictions were wrong, and all three err in the same direction — but they come from two different documents.** The ones wrong in A-6 itself are **two** (error rate 6~11% → 0.0% · defeater <50% → 85.4%), and the third (FOUND prediction 8~11 → 18) is in **the original pre-registration §7**. The third line of A-6 (`F_gen FOUND ≥ F_typed`) **was right**. We first attributed all three to A-6: the repair that fixed “selective reporting” **blurred the sources** in the same move. We correct it: **A-6 two + original PREREG §7 one = three sealed misses (two documents)**. The observation that the three are in the **same direction** stands — error rate below the prediction, FOUND above the prediction, defeater observability above the prediction.

**But writing “the three fold into a single uncalibrated instrument” was a logic error** (adversarial review R6). The three misses come from **different instruments**:

| Miss | Instrument | Explained by a “lenient scorer”? |
|---|---|---|
| Error rate 6~11% → **0.0%** | the local model adjudicates whether claims are true | yes |
| defeater <50% → **85.4%** | the **same** local model, a different task | yes |
| FOUND 8~11 → **18** | `found_keyword.py` — a **regex**. Does not pass through the scoring model | **no** |

So the precise form is this: **two fold into one explanation, the third does not.** The first two share the same ruler, so they must be counted as **one** piece of evidence rather than two independent ones (which is why “three independent pieces of evidence” was an over-count), and the FOUND miss remains outside that explanation and demands **its own check**. That check is the key-derivation control of §3.3 — keys were re-derived only from the benchmark’s clean control files and re-scored, with **0/72 units changed**. That is, the FOUND miss is **not** explained by “we fitted the keys to our results” (that control rules that explanation out).

Still, the threshold for the prescription is not lowered: this paper **does not use the error-rate point estimate as a basis for prescription until a positive control is applied to the F_gen arm** (§7 #9). The fact that there are three instruments only refutes “one explanation covers all”; it does not mean each instrument has been calibrated.

**The defeater axis — the number exists and the calibration does not.** A metric that was only in the pre-registration and had gone unreported was measured in this revision: the F_gen arm's defeaters **178/178** (17/17 units) were scored with the same local ruler → `IN-ARTIFACT` 152 · `NEEDS-OUTSIDE` 26 → **85.4%**. Per-unit values spread from **0%~100%** (h05 codex 0/3 · h04 codex 4/11 · g04 codex 27/27), so **the instrument is not stuck to one side.** But **no known-negative defeater was planted** — we did not insert a defeater condition "that we know to be unobservable" and check whether the instrument catches it, so **the level (85.4%) is `UNCALIBRATED`**. **Nor can the direction be used as it stands**: "a scorer leaning toward observable" produces exactly this miss. Therefore this miss is not evidence that "our defeaters were actually checkable" but **falls under the same single explanation as the two misses above.** To promote it to a claim about the world, a known-negative comes first.

**Correction (2026-09-12)**: this section as we first wrote it gave the adjudication rule as "three branches" (between / equal / equal-or-below), self-criticized that the measurement "fell outside the partition" and that "the pre-registered partition was incomplete", and further attached the post-hoc marker "the rule wrote 'equal' as an integer and we read it statistically". **All three disagree with the sealed text** — the seal is two outcomes and the word is "if similar". That narrative came not from the seal but from a restatement of the unsealed working results record, and in the process it **hit us harder than the record while at the same time writing the rule wrong.** We correct it.

**But the real weakness remains, and it is statistical power.** The difference this design (24 per arm, Fisher, α=.05 two-sided, baseline 0.58~0.75) can detect at 80% power is roughly **30%p or more**, whereas both contrasts of interest are **about 17%p** (14→18, 18→22). That is, **p=0.245 says little about the hypothesis** — "not separated" is a null obtained below the design's detection floor. That the seal's "if similar" branch fired is a fact, but **that "similarity" does not mean "no effect."** Looking at direction as well, the fact that the control's point estimate was in fact **higher** (22 > 18) is the stronger ground for the retraction.

**Incidental finding (not pre-registered)**: the blind arm's `unverified` jumped to 17/192 (F_gen 0/178, p=0.00001). The blind branch appears to have increased the places where "one's own family adjudicates one's own findings", and it is the main component that pulled coverage down to 65.6%. **Mechanism unconfirmed.**

---

# 6. A single observation on the decorrelation axis — family vs execution (this section is not subject to §5's retraction)

We ran two families on the same commit with the same prompt: codex 6S·3A (129k) vs Opus (**same family**, execution permitted) 1S·4A·7B (251k). Both sides carried findings the other did not, with the install reference-graph semantics (4 S findings) coming only from codex and the channel/format ones only from Opus. **S·A·B are the source record's own labels, and this paper does not define that scale.**

**The overlap/unique decomposition is not reported.** The working record's decomposition did not agree with its own totals (codex total 9 but 3+7=10; Opus total 5 but 3+3=6 — both exceeding by exactly 1 in the same direction), and telling which side is wrong requires the **finding-level lists**, which have not survived. Re-running the same task produces different findings, so that would be a different experiment rather than a recount, and this section's claim is about **that** run. We therefore report the totals and the observation below, and record the decomposition as **unrecoverable**. The observation stands without it.

Observation: giving the same task to different families left **unique findings on both sides**. The same family with execution attached was not 0 either, and the 4 install-surface S findings came only from the codex side.

**Do not read this as a "family effect" — it is a single observation** (adversarial review R8). Three things were not controlled: ⓐ the two executions have **different permissions** (only one was allowed to execute) ⓑ **different budgets** (129k vs 251k tokens) © **there is no within-family repeat baseline** — since we did not measure how many unique findings appear when the same family is run twice, how many of the 7+3 are "family" and how many are "run-to-run variance" **cannot be separated**. What §4.1 showed in two runs of the same design, 15/24 vs 20/24, is exactly the size of that variance. This section is therefore a record that "it occurred" and **not an estimate of an effect.**

**Do not confuse this with §5 — it is a different axis.** What survived here is "what you **send**" (the same task to different families), and what §5 retracted is "what you **show**" (another family's list in one family's round-2).

**Relation to external numbers**: [CG26](`arXiv:2609.10969`, 2026-09-10, 48 templates · 2,880 scenarios · fixed call budget) measured the two axes separately and reports that "cross-model voting reading the same evidence" approves 62.9% of risky proposals while "independent sources" approve 22.9%. In that paper's own words: *"The source effect is 40.9 percentage points, compared with 11.3 for model diversity."* **Citation caution — "3.6 times" is not that paper's claim.** 40.9 / 11.3 are their numbers, and the ratio (≈3.6) is **our computation**. They present no ratio (direct read of the abstract, confirmed 2026-09-12). Erasing this distinction becomes a citation mismatch.

**And we retract the further step we took here** (adversarial review R9). We first wrote *"§5's retraction points in the same direction"* — that is, the reading that our null supports their "the source axis is stronger". **The two are not the same manipulation**: what they manipulated is the **independence of evidence sources**, and what we manipulated is **whether the counterpart list is shown**. List disclosure does not change the source (both arms produced their own round-1 themselves). Using the null of a different construct as support for their result is direction-matching. This paper therefore **does not claim** that relation. The citation extends to "there is prior work that separated and measured this axis", and nothing is said about agreement with our numbers.

---

# 7. Threats and residuals (by name)

1. **The re-adjudicator is not a human** (the governor model — the second model that re-adjudicated a sample after the local scoring, §3.5). Hand-verified sample 40.
2. **h05 FOUND key UNCONFIRMED** — 1 unit of the FOUND figures rests on an estimated key.
3. **O extraction asymmetry** — octo prose was typed-extracted post hoc (a direction favorable to O, and as it turned out the O claims were real).
4. **reps=3 · 8 cases · arm-level repetition 1** — this project's own bar (reps≥3) is not met at the arm level.
5. **A single local scorer** — no family crossing. The bidirectional calibration table is **Appendix H** (Appendix K.20). That table's positive controls are **n=6 and all gemini·security outputs** — that is, the positive side of the calibration came from one family and one role only.
6. **F_r2ctrl prompt header wording differs** — structure identical.
7. **F_gen and F_r2ctrl do not share round-1** — r1 baselines 146 vs 136 (**7.4% difference**). The blinding applied only to round-2, but the executions differ from round-1 onward, so the growth-rate comparison (+21.9% vs +41.2%) is **not a one-variable comparison.** This is the **largest residual** keeping §5 bound to "not separated". **§4.1 closed this residual — without a new run, and in the direction unfavorable to us.** Round-1 and round-2 of the same run were already a "shared-r1 pairing". The closing result: round-2 contributes +3/+2 with **loss 0**, but the gains are concentrated in one case and are not significant, and the same-design round-1 baselines split **15/24 vs 20/24**. **Before that we tried to close it with a new run and did not run to completion**: a pairing arm sharing round-1 (Appendix F, seal `3b6b33cd…`) was designed, pre-registered and executed, but **both families' quotas were exhausted** and it stopped at **2/24** completed pairs. This does not mean that arm was unnecessary — n grows and window confounding shrinks. But **it is now not "without it we do not know the answer" but "it says the same answer in a narrower interval"**, and its priority drops below "increasing the number of cases". **§5's retraction stands.**
8. **The runs were scattered across several windows, and a selection rule sits on top of that** (Appendix K.21). The actual record: the 24 runs ran across **four windows** (a branch-switch accident **killed the usage of 18 runs** · codex quota exhausted 2 times · 1 ABORT from a leftover state file), and only the **"last successful run" per (case,rep)** was adopted as valid. Since that selection rule also applies to the 22/24 arm, it is a larger residual than "time window unmeasured" — there is no guarantee that the failed runs failed at random. The time-window effect itself remains **unmeasured**.
9. **Three of the five A-6 sealed predictions were wrong in the same direction** (§5.1). Their common explanation ("lenient scorer") **was tested with positive controls and weakened** (§3.4 — discrimination 12/12 · false rejections 0). Therefore **the cause of the three misses remains unknown** — the predictions may have been bad, the metric definitions may have differed from those at prediction time, or the arms may actually have done better than predicted. This experiment does not separate the three.
10. **Run-to-run variance is larger than the effect** (§4.1). Same-design round-1 at 15/24 and 20/24. The between-arm FOUND comparison does not separate in this sample. For the between-arm contrasts §4.2 **performed cluster correction** (Appendix K.22). **What remains uncorrected is the paired-gain side**: McNemar counts the 5 gains as independent, but those 5 are tied to **one case**.

11. **Defeater observability 85.4% is `UNCALIBRATED`** — no known-negative planted (§5.1).
12. **Pre-specification of the O-arm adjudication rule is unverified** — that document's last seal is 4 minutes 17 seconds later than the start of scoring (§4). The remaining five seals are in normal order.
13. **The cost of 2 passes was measured only in calls and wall-clock** — tokens and dollars are `UNMEASURED` (the external model CLI does not log them). No total is written (§8).
14. **The arm with the highest FOUND (F_r2ctrl) structurally has no cross-scoring contrast** — the places where that arm's claims overlap with the original scorers' (codex·gemini) verdicts number **0**, so it is "no contrast possible". That is, this arm's value hangs on the local ruler alone.
15. **We do not claim equivalence** — `p=0.346` (error rate F_gen vs O) has no power and **no non-inferiority margin was pre-specified.** It is not "the error rates are equal" but "this sample cannot see a difference".
16. **No paper-wide multiple-comparison correction was applied** — §4.2 applies Holm·Bonferroni only within the family of 4 contrasts (Appendix K.16). This paper carries 8 or more tests. At Bonferroni α≈0.006 neither of the raw Fisher heads (0.017 · 0.019) survives, and the cluster-corrected values (0.0625 · 0.109) are further still. **Therefore "significant" is not written at any level.**

---

# 8. Operational Harvest — How "Ran to Completion" Lies

F_pair could not produce a number, but it produced two results that anyone attempting to reproduce needs. Both lie on the same axis as the thesis of §1: **onto which surface an automated verdict can be placed.**

① **When the completion discriminator and the primary metric take different files as input, "ran to completion" lies.** Our execution chain counted directories and recorded "24/24 pairs ran to completion · scorable." The primary metric's input (`confirmed.jsonl`) existed for only **16/24** per arm, and the pairs for which the pre-registered roster actually held were **2/24**. The abort machinery was looking for round-2 members' failures under **round-1 filenames**, so 16 runs that died on a real quota passed as "not contaminated."

② **Do not hang validity on recognizing failure "vocabulary."** We missed three times in one day — a bare `429` matching inside `4294967295` in a finding's body, `Selected model is at capacity` absent from the vocabulary, one family's `Individual quota reached` absent from the vocabulary. The prescription is not to grow the vocabulary but to **switch to a positive condition**: "did the pre-registered roster all produce output?" Failure vocabulary is used only for the abort judgment "stop burning the window."

**There were five defects** (Appendix K.23). In addition to the two above:

- ④ **The preflight looked at only one side of the roster.** There were two families but only codex was checked, and so **stage 2 ran to completion with gemini dead.** If the roster has n members, the preflight checks n — the same principle as the "positive condition."
- ⑤ **The error channel was in fact a 254KB transcript** (including the prompt echo). So **the target source's own log string** (`logger.warning("Rate limit exceeded, …")`, `case_g04.py:1608`) matched as "our quota." The instrument's input was not "errors" but "everything," and this is not a vocabulary problem but a **channel-definition problem**.

**③ When the model spends its entire budget on the "thinking" field, the response is an empty string and the exit code is 0.** In defeater-axis scoring, of 17 units, 2 were recorded as `scored=0` — rc=0, `eval_count` 8,225 (a plausible value), 0 errors. The artifact showed `done_reason=length`, the content was **entirely in the `thinking` field**, and `response` was 0 bytes. **The possibility that the missingness correlated with unit size is circumstantial at best** — the two that died were def=**35** (of 17 units, rank 1) and def=**13** (tied for 5~6th), while **27 · 19 · 18 are larger and did not die.** With n=2 one cannot speak of correlation either (Appendix K.24). **But the evidence that the bias was real is not the size correlation but the values themselves**: **100% IN-ARTIFACT** in the pre-truncation sample, **85.4%** in the full set. Aggregated as-is, we would have reported an instrument artifact as a result. Our interim aggregate actually rode that bias (100% IN-ARTIFACT in the pre-truncation sample, 85.4% in the full set). The prescription is neither vocabulary nor retry but **reading the completion reason**: `done_reason=length` → write no answer, record it labeled `TRUNCATED`, and size the output budget to include the thinking share. **fail-before/after remain in the same log** — `ctx=16384 → scored=0` · `ctx=24576 → scored=13 done_reason=stop`. **The part more useful to a reproducer is this — the defense already existed in the neighboring file.** The main scorer (`score_local.sh`) had suffered the same accident earlier and already had a complete defense (separate output budget · `n==0` as `UNSCORED_NO_CONTRACT` · partial scoring as `TRUNCATED` · **seeds placed last in the list to surface truncation**). The defeater scorer written later **did not inherit that defense.** So the class is not "did not read the completion reason" but **"two call sites in the same role diverged in leniency,"** and that does not close by following the discipline more faithfully — it closes by consolidating the call sites into one. **The numbers in this paper are unaffected by this defect — verified exhaustively**: the four arms' scoring logs total **58 units, all `status=OK`** (F_gen 17 · F_typed 15 · F_r2ctrl 18 · O 8), and there are **0** empty output files. Not one truncation or partial score.

**⑥ 2-pass cost — we retract our first conclusion on this item. It does not exceed the prediction.** Pre-registration §7 sealed "1.5~2.0× **F** (2nd pass = 1 additional call per family)" (Appendix K.25).

We first wrote *"correct in calls, exceeding at 2.2~2.4× in wall-clock."* **That wall-clock value was wrong, and wrong in two layers** (adversarial review R10):

- **E1 estimator** — 2.2~2.4 is `(per-stage median r1 + per-stage median r2) / per-stage median r1`. **Medians are not additive.** On the same data, the median of per-run total time (315.4 s) and the sum of stage medians (296.8 s) differ by 6%.
- **E2 denominator — the larger error. The runner had already been recording per-run total wall-clock** (`run_arm.sh:33` measures with `date +%s` and writes

wall=Ns to runs.tsv). The instrument (cost_2pass.py) **never read that file** and re-estimated only the fleet interval from file mtimes. Outside the fleet window there is pipeline time with a per-run median of 121~135 s. Using a partial interval as the denominator and the whole of stage 2 as the numerator inflates the multiplier. And mixing the two in one sentence meant **the two multipliers in the same sentence used different denominators** — calls relative to F_typed, wall-clock relative to the arm's own round-1. The prediction's denominator is "F."

**Re-measured with the canonical instrument** — per-run total wall-clock from runs.tsv, paired with the 1-pass arm F_typed by case__rep, the 18 runs that died with rc=2 excluded (all n=24 · 0 missing):

| Contrast | Ratio of medians | Median of ratios | Ratio of means |
|---|---|---|---|
| F_gen / F_typed | **1.60** [1.49, 1.99] | **1.97** [1.73, 2.19] | **1.86** [1.62, 2.12] |
| F_r2ctrl / F_typed | **1.80** [1.64, 2.15] | **2.04** [1.62, 2.53] | **1.89** [1.61, 2.20] |

**Why all three are reported**: ① and ② diverge by up to 0.37 (1.60 vs 1.97). The distribution of ratios has a long right tail (0.79~3.78), so choosing any one of them would itself be a selection.

**Cross-check by an independent route**: within the same arm, subtracting "had this run been 1-pass" and re-measuring gives 1.58~1.79, and the 1-pass median it implies, **254 s**, meets the actual 1-pass arm's **267 s** within 5%.

Therefore **calls are exactly 2.00×** (relative to F_typed) and **wall-clock is 1.6~2.0×** (relative to F_typed) — **not exceeding the pre-registered prediction of 1.5~2.0×.** Of the six estimators, only one exceeds 2.0 (F_r2ctrl median of ratios 2.04, CI lower bound 1.62).

**And the "cause" we attached is also refuted.** We wrote "the counterpart's list is attached, so the prompt grows," but the control arm that **hid** the list had a prompt only **1.05×** larger and its round-2 was still slower (the disclosed arm: 1.21×). Prompt size does not explain it. Output volume is a candidate (disclosed 1.13× vs hidden **1.67×**) but only as far as correlation — **cause undetermined.**

**Tokens and dollars remain UNMEASURED**: the external model CLIs do not leave them in the log. **We write no total. Four remaining limits, by name.** ⓐ Per-round "canonical" time is **not computable**: runs.tsv records only total time and has no per-round timer, so within-arm pairing rests on the **unverified assumption** that "non-fleet overhead is independent of pass count." ⓑ F_typed differs not only in pass count but **also in prompt contract**, so there is no clean pass-count contrast; the evidence is that **two imperfect baselines agree**. ⓒ The window effect is unresolved (§7 #8) — each arm has 1 run in which "2 passes were faster." ⓓ Sensitivity to duplicate-key handling (a first-line rule flips it back to 2.4~2.7) is closed by a **discriminator**, not by convention: the disk-mtime interval is a lower bound on wall time, so of the 26 duplicate keys, **last line only is consistent for 24 · both possible for 2 · first line only for 0**.

**So this item is not a "sealed miss"** — what missed was our instrument. We first placed it beside the three misses of §5.1 as "a fourth miss in the other direction," and **that placement itself was wrong.** The prediction was right and the measurement was wrong.

**Budget numbers obtained incidentally — written as three separate things** (Appendix K.26):

- **Measured**: calls actually made by one bucket of one family, **31** · the other family, **45**.
- **Derived (not measured)**: **72 per family** needed for 1 pairing · **56** for the resume plan. Values computed from the design.
- **And those numbers are gradations, not ceilings**: that family's quota screen reads "Quota is consumed proportionally to the **cost of the tokens**" — consumption is proportional to **cost**, not to the "number" of calls. So 31/45/56/72 must not be read as "how many times we can call."

**Two framings are retracted** (Appendix K.26); the corrected forms are that that family's reset is **weekly**, and that the right conclusion is "**even one weekly reset may not suffice**." And **"about 61 hours between the two windows" is not an observation**: that number is an estimate of "time remaining until reset" at some moment, later replaced by the CLI's direct reading `Refreshes in 58h 48m`, and **since the resume itself has not yet happened, no observation of a window interval exists.** The planning unit is not "days" but **"that family's reset cycle."**

**There was no data fabrication.** Failed runs were honestly recorded as `UNREVIEWED` and produced no verification output, and the pipeline guard recorded at every run that "if one family dies, cross-verification does not hold." What collapsed was **the aggregation layer above it.**

---

# 9. Conclusion

We write the claim that survives on the same ruler in two layers. This section records **what stands and what does not**; how those claims narrowed, and from which draft, is written item by item in **Appendix K.27**.

**What stands — three of them, and all three are observations, not guarantees** (Appendix K.27):

- **Observation**: in within-run pairing, **the number of units whose planted-defect FOUND state flipped 1→0 is 0** (0/24 · 0/24). The conditional denominator is not 24 but **15 · 20**, the units already FOUND in round-1.
- **What that observation does not say.** ① FOUND is a **binary indicator** of "was the planted defect pointed at," so preserving its state does not mean **individual claims were preserved**: other valid findings, or the accuracy of explanations, may vanish where this indicator cannot see them. ② That the **total count** of claims rose is not evidence of individual preservation either. ③ The two arms use **the same 8 cases**, so this is not 2 independent replications. ④ It implies nothing at all about future loss or **non-inferiority** (no permitted loss margin was pre-specified).
- **Error rate**: point estimate 0.0% on the local ruler. **The 95% upper bound ≈2.1% (rule of three) is a reference interval under the "independent trials · complete adjudication" assumption** — for the same paper to acknowledge that claims are bundled within cases and then use an independent binomial interval is inconsistent, so until clustering is reflected we **do not cite it as an upper bound on the actual error rate.**

**What stands goes as far as "in this sample we did not observe a loss of planted-defect detection."** So we also do not write **"pays the cost but takes no penalty."**

**What does not stand**: "look twice and catch twice as much." There are three reasons, mutually independent — ⓐ the between-arm FOUND gap **does not separate from run-to-run variance** (identical-design round-1 at 15/24 vs 20/24, §4.1) ⓑ the paired gain is not significant and is **entirely in one case** (§4.1) ⓒ **after cluster correction not one contrast is significant** (§4.2 — 0.017·0.019 → 0.0625·0.109). The cause first named for that gain, "disclosure of the counterpart family's list," also did not separate from the control and was **retracted** (§5).

So the prescription also narrows: **"do not sell on the basis of how much it catches"** is as far as this data goes, and we **cannot yet sell** on the basis of zero loss (①~④ above). And to answer "how much more does it catch," **increasing the case count comes first** — that the gain was tied to a single case points that way. **But do not read this as "more reps are not worth it"** — in cases whose success probability is extreme, unanimity is common even under conditionally independent repetition, so with the present table the value of additional reps **cannot be adjudicated** (Appendix K.27).

**And why the error rate is cited**: that ruler's discrimination was pre-registered and measured — fabricated claims built at the arm's own difficulty were **rejected 12/12** while real claims were retained 11/12 (0 false rejections, §3.4). The prediction was 40~75% and was wrong. Since n=12 we write "point estimate 100%, 95% lower bound 75%." In exchange, **the cause of three of the five sealed predictions being wrong remains unknown** (§5.1).

**Finally, what this paper says about its own method — the free lunches were three.**

| Metric | What can distort it | Evidence grade in this paper |
|---|---|---|
| Precision | **Selective deletion** — the error rate of what was dropped never appears in the report | **Motivating case** (the 8→6 anecdote of §1 — no denominator, no test) |
| "Ran to completion" | **Choice of discriminator** — count directories and a dead run has run to completion | **Observation** (§8 ① — recorded as 24/24, actually 2/24) |
| **Significance** | **Independence assumption** — count bundled reps as independent units and p goes down | **Observation** (§4.2 — 0.017 → 0.0625) |

**Separating the evidence grades is the point of this table** — the first row is what §1 itself called an "anecdote," and only the lower two are observations of this study. Nor is it true that "the independence assumption always lowers p" — it did so on this data. So the prescriptions are also three: make the drop audit a completion condition, change the completion discriminator to the primary metric's **input**, and do not erase the uncorrected p but **write the corrected value in the adjacent column.** The third was newly added in this revision, and it is what brought this paper's significance claim down entirely — not erasing that fact is the point of this section.

---

## 10. This Paper's Contributions by Name — and What "Going Beyond octo" Means

The primary result of this paper is **null**. So unless it states its own contributions, it reads only as "a paper that failed to show." We separate them into §10.1~§10.4 (and its sub-section §10.4.1), and **wrote only what can be verified inside this paper** — no general propositions borrowed from outside are here. (§10.5 is not a contribution but **what closes it**.)

**But these are not of equal rank. We write the grades first — the prior-art search narrowed this section six times, and what remained after narrowing is the thesis.**

| | Grade | Why |
|---|---|---|
| **§10.1** We built the measuring ruler first and killed our own main effect with it | **Contribution** | Verifiable inside this paper, and an axis for which the prior-art search found no precursor. What gets reported is "what was found"; "which check was fixed before seeing results" is not reported |
| **§10.4.1** $\hat{\varepsilon} = \hat{\varepsilon}(\pi)$ — subset selection redefines the **denominator** of the reported metric | **Contribution** | The prior lineage goes as far as "reporting is distorted" and does not write the denominator redefinition. But **the incentive clause is not ours** (§10.4.1 table); what remains is only that identity and **the machinery that enforces it** |
| **§10.2** Loss 0 vs gain | **Observation** | Not raised to a contribution — **variance (≈5) exceeds the effect (3, 2)** and no design measuring dispersion as the primary metric was included |
| **§10.3** A decomposable number | **Property statement** | A property of the pipeline, not a finding of this paper. The content goes as far as the **fact** that the decomposition is possible in one direction only |
| **§10.4** The number does not buy a surface | **Rule** | A rule this paper **keeps**, not a proposition this paper **established**. The cost-structure rationale we first wrote is retracted (below), the asymmetry sub-argument has [OCR26] as prior, and placing a human at the terminal is itself done by [EBP18] |

**Therefore what this paper claims as "ours" is only the two rows above.** The other three are included but not raised to contributions. The reason the narrowing history is not erased is §10.1 itself — if which claim was abandoned when is not preserved, §10.1 cannot keep its own instruction.

## 10.1 We Built the Measuring Ruler First, and Killed Our Own Hypothesis With It

The order is the contribution. We measured the scorer's discrimination with **pre-registered positive controls** (§3.4, 12/12 rejected · 0 false rejections), ruled out the suspicion that the FOUND key had been fitted to results by **re-deriving the key from a clean control file** (§3.3, 0/72 changed), scored the defeaters exhaustively (§5.1, 178/178) — and **when we measured our own main effect with that ruler, nothing remained** (§4.2, 0/4 after correction). And **we did not erase the raw p but wrote the corrected value in the adjacent column.**

Reverse this order and the same data yields a different paper: one that reports 0.017·0.019 without correction, does not run the positive controls, and uses "error rate 0.0%" as the basis of a prescription. **That paper too could have been written from this data.** The difference between the two papers is not a finding but **which check was fixed before seeing results**, and that is the axis that goes unreported in this field.

## 10.2 A Harness's Value Is Not "Exceeding the Ceiling" but "Stably Extracting Up to the Ceiling"

That is the quantity this paper actually measured ("harness" here is [ZWF26]'s sense as cited in §2.5, and the pipeline of §3.1 is that layer). **What stands is loss 0** (§4.1) — the second pass did not subtract from the first pass's planted-defect detection (FOUND). **What does not stand is the gain** (§4.2). That is, the property this pipeline showed is not "makes the model do what it could not" but rather "loses less of what it was doing."

**And that distinction touches this paper's largest residual**: two runs of the same design split at 15/24 and 20/24 (§4.1). **Variance (≈5) exceeds the effect (3, 2).** So the most accurate sentence that can be said about this pipeline now is — **not even "it reduced variance," yet.** If reducing variance is the goal, it must be **measured directly** (dispersion over n repetitions of the same design), and this study did not put that measurement in its design. What can be established here goes as far as "we observed that variance exceeds the effect," and that **changes the primary metric of the next experiment** — it must be **dispersion**, not mean harvest.

## 10.3 "Going Beyond octo" Is Not a Number but Decomposability

This paper **does not claim** a FOUND advantage over octo (corrected p=0.109). Then what remains — **the fact that the decomposition is possible in one direction only** remains.

On the coverage axis **there is no between-arm comparison to make** — the two sides' figures come from different instruments, and that axis's test is retracted (§4). What exists is a decomposition on one side only: of our 35 losses, **24 are "outside file scope"**. The same decomposition cannot be applied on octo's side — **because that record does not exist.** So what is one-sided here is not a ranking but **a record**: on our side "why" is answerable, on octo's it is not.

Therefore what this paper offers is **not a larger number but a decomposable number**. And decomposability is not a virtue of the paper but **a property of the pipeline** — produced by a design that carries claims typed, counts drops by reason, and does not fold unmeasured into 0. **Without that property the self-refutation of §10.1 above is impossible too** — without knowing what came from where there is no means to refute one's own result. This is the whole of the "going beyond" this paper speaks of, and **it is not a performance claim.**

## 10.4 The Number Does Not Buy a Surface — the Rule This Paper Keeps

The claim error rate point estimates are 0.0% (ours) and 1.1% (octo). **However good that number becomes, this paper does not prescribe placing it on the "delete · publish · history-rewrite" surfaces.**

**We retract the rationale we first wrote here (2026-09-17).** That sentence was *"what separates the two surfaces is not the size of the error rate but the **cost structure** when wrong, so improving the number cannot promote a surface."* **Under finite cost that inference is false** — even if the cost of a wrong publish is 10,000× that of a wrong review, lowering the error rate to 1/10,000 or below reaches the same expected loss, so the conclusion becomes not "promotion impossible" but "a stricter threshold needed." **Irreversible is not infinite** — even deletion is modeled as finite loss depending on backup, audit-log, compensation and rollback design. So that sentence depended on an **unstated premise** that "cost is infinite." (Pointed out in adversarial review, and the authors did not find it on their own. We record that fact at the same grade as the operational harvest of §8. Appendix K.28.)

**The replacement rationale claims nothing about the size of cost at all.** It has three legs, and the first leg is not ours:

- **L1 Non-transferability (external)** — distribution-free risk certification does not hold when deployment changes. Kotte [K26] measured that under cross-dataset shift, of 16 transfers, the target is **violated in 14** (static CRC and every ACI step size tested), that of the 16 a valid anytime-valid monitor **certifies 0**, and that practical certification opens only at **α = 0.40**. So even the "strict version" of the number **does not carry** across surfaces and deployments.
- **L2 Self-measurement + purchasability** — that error rate is produced by **the very system family that would receive passage, measuring itself**, and it can be lowered **by discarding claims** without fixing behavior ($\hat{\varepsilon} = \hat{\varepsilon}(\pi)$ of §2.4). The incentive clause itself is prior — Goodhart [G75, ME25] · NIST `MEASURE 1.3` [N23] · and the impossibility of a threshold hung on a self-reported number was proven **as a theorem** by Lovén et al. [LDM26]. Ours is only the identity "subset selection redefines the denominator of the reported metric" and **the machinery that enforces it**.

- **L3 Authority (lexicographic)** — that the terminal responsible party must be **a human or a recorded override** is a hard constraint **above** expected utility. This leg is not ours alone either — even error-budget policy **escalates to a human** on disagreement [EBP18]. **The precise difference is not "whether a human is placed" but "whether the number opens the gate"**: an error budget keeps deploying while budget remains; with us, the number does not open the gate.

Therefore **what this section claims is only this: $\hat{\varepsilon}$ does not identify the quantity the gate needs, so there is nothing to hang a threshold on in the first place.** It says nothing about the size of cost.

**And the asymmetry argument of this section is not our finding.** "Expensive errors are silent errors" is already in the shipped filter prompt of a production code-review tool [OCR26] — *"Keeping an incorrect comment costs a reviewer a few seconds of attention. Removing a correct comment **silently destroys a real finding. It never reaches anyone, and nobody learns that it was dropped."*** But that side separates FP from FN **within one surface** and biases policy toward "keep," while we separate **between surfaces**. The shared sub-argument is handled by citation, and only the between-surface application is written as this paper's (§2.3).

There are two places where this rule actually binds in this paper: ⓐ our pipeline does not perform "candidate deletion" at the verdict stage but **counts drops by reason** (§3.1) — the route of raising precision by deletion was blocked in the design ⓑ the error-rate point estimate of 0.0% is **not used as the basis of a prescription** (§7 #9, until positive controls are run on the F_gen arm). ⓑ is a constraint that works against us, and so that place shows this rule is not decoration.

### 10.4.1 Formalization — $\hat{\varepsilon}$ Does Not Identify $R$

§10.4 above says in prose that "the number does not open the gate." This sub-section writes, as an identity, **what makes that sentence hold**. **First we separate what is not claimed — the eight rows below are all prior and are handled by citation only.** Without this table the following paragraphs would be uncited prior art.

| Claim this section does not make | Prior |
|---|---|
| A threshold hung on a self-reported number is not incentive-compatible | [LDM26] — theorem |
| Calibration distribution ≠ deployment distribution | [PZM20] |
| Distribution-free guarantees break when deployment changes | [K26] — of 16, 14 violated |
| Observed performance is not grounds for authority (half) | [A26] |
| When a measure becomes a target it ceases to be a measure | [G75] · formalization [ME25] |
| The maker must not grade itself | [N23] `MEASURE 1.3` |
| When the audited party controls what is measured, the metric is manipulable | [BD26] — impossibility |
| The evaluated party moves to obtain a favorable verdict | [HMP16] |

**Setup.** A verdict pipeline produces claims over a set of items $X$. For each item, $e(x) \in \{0,1\}$ denotes whether the claim about that item is wrong, and $m(x) \in \{0,1\}$ whether a real defect exists that was not claimed. $\pi : X \to \{\text{keep}, \text{drop}\}$ is the **retention policy** and $K(\pi) = \{ x : \pi(x) = \text{keep} \}$. The reported error rate and the quantity the gate actually needs are, respectively,

$$\hat{\varepsilon}(\pi) = |\{ x \in K(\pi) : e(x) = 1 \}| \,/\, |K(\pi)|$$

$$R(\pi) = c_{wrong} \cdot \Sigma_{x \in K(\pi)}\, e(x) + c_{miss} \cdot \Sigma_{x \notin K(\pi)}\, m(x)$$

**The definition of $\hat{\varepsilon}$ has no second term. This sub-section is the consequence of that one line.**

**P1 (purchasability).** If $K(\pi)$ contains even one item with $e(x)=1$, then for the $\pi'$ that excludes only that item, $\hat{\varepsilon}(\pi') < \hat{\varepsilon}(\pi)$ and **the generator does not change by a single bit.** It holds under a weaker condition too — dropping the top $k$ by any signal positively correlated with $e$ lowers $\hat{\varepsilon}$ in expectation, and **the correlation need not be strong. This proposition is nearly trivial and we do not hide that.** Its worth lies not in the depth of a theorem but in that this identity specifies "what the machinery must demand" (§10.4 ⓐ·ⓑ).

**P2 (non-identification).** $\hat{\varepsilon}(\pi)$ is an estimator of $E[e \mid \text{keep}]$, and $R(\pi)$ adds $\Sigma_{x \notin K(\pi)}\, m(x)$ on top of it. **$\hat{\varepsilon}$ is silent about that term.** So two pipelines with the same $\hat{\varepsilon}$ can diverge arbitrarily in $R$ — one may have dropped nothing, the other may have dropped the hard half. Therefore **unless the drop set is audited, $\hat{\varepsilon}$ is not a proxy for $R$.**

**P3 (interest binding).** $\pi$ is set by the party seeking to pass the gate, and by P1 moving in that direction is **always possible, and $\hat{\varepsilon}$ itself retains no trace of it. The incentive clause is not ours** (rows 1·5·6·8 of the table above). **Ours is where it joins P2 — "the trace does not remain in $\hat{\varepsilon}$"** — the prior lineage goes as far as "reporting is distorted," and none of it writes "subset selection redefines the **denominator** of the reported metric." Even the closest, [A26]'s *"corrected performance can hide the competence signal"*, has **"downstream correction" as the erasing agent, not "the evaluated system's own $\pi$."**

**Why we use no cost — and why that saves this section.** The road of arguing by cost structure using $c_{wrong} \cdot c_{miss}$ is closed: Franc, Prusa & Voracek [FPV23] showed **as a theorem the equivalence** of cost-based, bounded-improvement and bounded-abstention with *"despite their different formulations the three rejection models lead to the same prediction strategy: the Bayes classifier endowed with a randomized Bayes selection function"* — changing the formulation by itself yields a gain of 0. **P1-P3 are not caught by that theorem**: [FPV23] **assumes the true risk is known** and discusses the optimal strategy, while this section attacks "what stands in for that true risk," so the layers differ. Therefore the conclusion of §10.4 is not "a stricter threshold" but **"there is no object to hang a threshold on at all."**

**Five named limits.** ① **P1 says nothing about an upper bound** — "manipulable" and "how much was actually manipulated" are different quantities, and the latter is not in this section. ② **It assumes $e(x)$ exists** — in our setup $e$ is a value set by yet another adjudicator, so **there is recursion**, and in the field measurement that adjudicator had to be calibrated with seed controls (§3.4). ③ **It is not a theorem** — P1 is nearly trivial and P2 is a consequence of the definition, so we **do not present it at the same grade** as the impossibility of [LDM26]. ④ **No cost parameters are used** (the paragraph above). ⑤ **It inherits the limits of the field evidence** — organization n=1 · adoption unmeasured · the **number of cases** in which this gate actually prevented an incident **untallied**.

**Falsification conditions.** ⓐ If some prior work has already formalized **that subset selection redefines the denominator of the reported metric**, this sub-section dissolves — **the current strength of "none" goes as far as "none at the title·abstract layer"** (the list of unopened prior work compared only at the title·abstract layer is not included in this submission). ⓑ If a case is constructed in which $\hat{\varepsilon}$ fails to identify R even with drops audited, ⓐ of §10.4 collapses and reversible surfaces too must move toward "not reading the number." © If someone proposes an **error-rate threshold on an irreversible surface** on the basis of this section and we cannot produce a sentence that blocks it, that is a failure of the second rule this paper said it keeps.

## 10.5 So What Closes It — Now Writable as Numbers

| Open | What is needed | Scale |
|---|---|---|
| "Does looking twice catch more" | **Case count** (not reps — §4.3) | assuming a true win rate of 0.6, **12 cases**; at 0.4, 19; at 0.3, 25 |
| Does it reduce dispersion (§10.2) | **n repetitions** of the same design as the primary metric | Undesigned — replace the next experiment's primary metric |
| Individual claim preservation (§4.4) | **Claim records** (stable id · provenance) in the pipeline | Structurally impossible with current data — new runs |
| Transfer of positive controls (R7) | A corpus of **naturally occurring errors** | Only synthetic mutations exist — new corpus |

**The three below are listed as this paper's "limits" and are not closed.** Closing them means a new experiment, and that experiment must be designed to close **three of the four above together, not one** (a run with claim records, using a natural-error corpus, closes R1·R7·R11 at once). Putting that design in the present paper would be **a promise, not a result**, so here we write only scale and conditions.

## 10.6 The Practice This Method Belongs To — and What It Does Not Claim

What this paper reports is the methodology layer of **our way of AI governance engineering** — *governing the error rate of a process that produces judgments, by grade, according to the reversibility of the surface that judgment is used on*.

**Neither the name nor the field is ours, and this section makes no such claim.** The enforcement layer has several precedents. A policy-enforcement engine inside a harness already uses our vocabulary in our sense — the first sentence of [ZWF26]'s abstract reads *"AI agents increasingly run in production through harnesses, the* *software around the LLM, including an engine that enforces safety and effectiveness policies, e.g., 'run tests before committing.'"* Three further lines exist. [OC26] builds a governance gate that opens and closes the scope of automated enforcement on signal thresholds — *"A Governance Gate built on these signals achieves 78.6% automation coverage with 64.9% risk reduction."* [BSN26] gives an architecture that accounts for residual violations by the error rate of the enforcement stack **itself**: residual hard violations "scale with enforcement-stack error rather than environmental violation opportunity." [PN26] translates norms into executable compliance pipelines, with 43 machine-checkable acceptance criteria across six modules. The organizational layer already has its implementation-and-difficulty study in [MF22]. **What this paper claims is neither the name nor the field, but one implementation and the record of how it broke.**

**How that practice relates to this paper.** The first of the three paths in §8 — precision improving for free through selective deletion — is what that implementation calls *judgment-channel integrity*. If the error rate of the discarded claims never reaches the report, the remaining number cannot be interpreted; so the implementation **refuses to complete** a run whose drops were never audited. This paper names that place, records that **we ourselves took it** (§8 ①), and carries the prescription of making a drop audit a completion condition — §10.1's "we fixed the ruler first, and the ruler killed our own hypothesis" is that discipline operating inside this paper. The doctrine layer itself is treated in [Kw26].

**The failure of step-level compliance to compose has recently been given a name.** [KGG26] defines it as *"the failure of step-level compliance to compose"* and gives a taxonomy of four types: Authority Creep, Threshold Laundering, Cumulative Sum Violation, and Context Collapse. The surface-grade rule in §10.4 governs judgments **within one surface**, so composition across multiple steps is a place this paper does not address.

---

# Appendices

**The scope of the supplement, first.** The pre-registrations and the instrument scripts are **not attached to this posting** — arXiv's PDF submission path requires the submission to be exactly one PDF file (measured 2026-09-18: bundling them with the PDF, and uploading them separately after the PDF was accepted, are both refused, and for the same reason). **The supplement manifest carries the sha256 and md5 of all 41 files instead, and the files are available on request.** A reader cannot check them against it today, but any copy released later must match those fingerprints, so the claim stays falsifiable.

**Three of those files are not released even on request**: Appendix C (`PREREG_ADDENDUM3`) and two scoring scripts (`score_local.sh`, `score_defeater.sh`). All three carry the **private-network address** of the host the local scoring model ran on, which cannot go onto a public surface. Their fingerprints are in the manifest as well, so the same condition binds them. This restriction is the decision recorded in this paragraph, not an omission.

This paper does not delegate numerical authority to working records the reader cannot inspect. The pre-registration appendices (A~G) each have a sealed file (table below — available on request), and the body's figures are presented so that they close inside the body, together with their definitions and recomputation (Appendix K.29).

Appendices H·I·J are "result tables", which is a different thing from the "pre-registrations" of Appendices A~G — calling them by the same name makes "already attached" and "absent" indistinguishable. In particular, Appendix E is the **pre-registration** of the governor re-adjudication and Appendix J is the **result table** of that re-adjudication (Appendix K.30).

Appendices A·F·G carry seal hashes; H·I·J cannot — a result table is an output, not a pre-registration, so the property "fixed before the results were seen" does not exist for it in the first place. The reproduction path for those three is not a seal but the **instrument list** (below).

**Appendices A~G — pre-registration documents (available on request).** Each appendix is one sealed file, and the body refers to it by its seal hash.

## Appendix A — arm definitions · A-6 sealed predictions (5 lines) · A-7 revert to the original plan

Sealed file: `PREREG_ADDENDUM_2026-09-10.md` (`19bbff9f…`) — available on request.

## Appendix B — scoring routing

Sealed file: `PREREG_ADDENDUM2_2026-09-10.md` — available on request.

## Appendix C — scorer calibration

Sealed file: `PREREG_ADDENDUM3_2026-09-10.md` — **not released** (private-network address — see "The scope of the supplement" above). Its fingerprint is in the supplement manifest.

## Appendix D — O verdict rules

Sealed file: `PREREG_ADDENDUM4_2026-09-12.md` — **seal time UNVERIFIED** (§4) — available on request.

## Appendix E — governor re-adjudication, 40 items (pre-registration)

Sealed file: `PREREG_ADDENDUM5_2026-09-12.md` — result table in Appendix J — available on request.

## Appendix F — paired arm (did not run to completion, §7 #7)

Sealed file: `PREREG_ADDENDUM6_2026-09-12.md` (3b6b33cd…) + NOTE — available on request.

## Appendix G — F_gen positive control

Sealed file: `PREREG_ADDENDUM7_2026-09-12.md` (b6b1d65c…) — available on request.

Full seal hash of Appendix F:
`3b6b33cd03064f9bf7cdd92f7e12a5d398bf60fcf778ddaac46b5bee15644183`
(`PREREG_ADDENDUM6_2026-09-12.md`) — accompanied by the NOTE written after the run (`PREREG_ADDENDUM6_NOTE_2026-09-12.md`).

## Reproduction path — instrument list

The directory's **33 instrument FILES** (lanes excluded) are listed in three groups (Appendix K.31). **The "22 kinds" in K.31 and this 33 count different units** — 33 is files, 22 is kinds (several files belong to one kind).

- Instruments: `found_keyword.py` · `score_local.sh` · `score_tally.py` · `quota_check.py` · `pair_validity.sh` · `found_within.py` · `paired.py` · `prompt_parity.sh` · `score_defeater.sh` · `cost_2pass.py` · **`seal_audit.sh`** (seal content integrity + temporal order)
- The eight that produce the cited numbers: `cluster_test.py` (§4.2 permutation · sign p) · `multiplicity.py` (§4.2 Holm/Bonferroni) · `power_analysis.py` (§4.3 power · realized type I error) · `claim_correspondence.py` (§4.4 retention interval) · `found_keyword2.py` (**0/72** derived-key re-scoring) · `build_poscontrol.py` · `score_poscontrol_tally.py` (**12/12 · false rejections 0**) · `seed_label_probe.py` (label-reading rejected 12/12).
- The remaining 14 are the input-generation and execution chain: `build_score_input.py` · `build_O_score_input.py` · `seed_score_input.py` · `score_run.sh` · `score_run_seeded.sh` · `run_arm.sh` · `run_pair.sh` · `pair_chain.sh` · `codex_preflight.sh` · `score_F_r2ctrl_chain{,4,5,6}.sh` · `scan_r2_loss.py`.
- Lanes: `test_quota_check_lanes.sh` (10) · `test_pair_validity_lanes.sh` (7)

## Appendix H — bidirectional calibration table (three scorers, same ruler)

| Scorer | Positive (hand-verified REAL → REAL) | Negative (seed → NOT-REAL) | Disposition |
|---|---|---|---|
| codex | **0/6** (all UNVER) | 24/24 | **Negative side only calibrated.** Abstention rate 62.7% — therefore its verdicts are reported only as an "abstention rate" |

| Scorer | Positive (hand-verified REAL → REAL) | Negative (seed → NOT-REAL) | Disposition |
|---|---|---|---|
| gemini | 6/6 — **self-scoring probe** | 28/28 cross + 12/12 self | Counted as calibrated only on the cross-routed half (the 70 codex-produced items) |
| local (3rd family) | **6/6** | **56/56** | **Bidirectionally calibrated.** Therefore the scorer whose error rate may be cited |

**gemini's positive 6/6 is not used in computing the error rate** — being its own claims, the direction leans toward granting REAL, so it is **upper-bound evidence** only, and the pre-registration (Appendix B §5) declared it as a control. **Named limitation — the same as §7 #5, restated here**: positive controls **n=6 (3 claims)**, **all gemini-produced, security role**. That is, the positive side of the calibration came from **one family and one role** only. Auto-generated positive seeds remain unresolved, since they require producing "true defects" mechanically; the realistic path is to add hand-verified anchors.

## Appendix I — re-scoring of O (octo) on the same ruler

Same ruler = local qwen3.8:27b · temp 0 · no answer key · restricted to the claim list · 3-tier seeds placed last. O's claims were **extracted post hoc from the prose** of the 24 O_nosent runs (94 items, 7 empty runs). **That extraction is a judgment, whereas F_gen is pipeline-typed — the asymmetry runs in O's favor** (ambiguous sentences are "counted in" as claims).

| | F_gen | O_nosent (re-scored) |
|---|---|---|
| Claims | 178 | 94 |
| Error rate (NOT-REAL/total) | **0.0%** (0/178) | **1.1%** (1/94) |
| Non-REAL (NOT-REAL+UNVER) | 0.6% | 8.5% |
| Seed capture | 60/60 · 17 units CALIBRATED | 28/28 · 8 units CALIBRATED |
| FOUND (same keyword instrument) | **18/24** | 9/24 |

```
error rate 0/178 vs 1/94       Fisher p = 0.346   not significant (§7
#15: equivalence is not claimed either)
FOUND   18/24 vs 9/24       Fisher p = 0.019   after cluster
correction 0.109 — §4.2
```

**The coverage row and its `p<0.001` were deliberately omitted from this table.** The original record carries that row, but body §4 **retracted** it (the value compared F_gen's "verification pass rate" 122/178 with O's "adjudication rate" 87/94, and the two numbers are different instruments). If an appendix carries a number staler than the body, that spot becomes the source of the next correction, so **the retracted row is** **not carried over. The sealed prediction was wrong** — it predicted "local error rate_O ≥ 2.7%" and 1.1% came out, in the direction **favorable to O**. The pre-fixed Pareto condition (error rate_F ≤ error rate_O ∧ FOUND_F ≥ FOUND_O) is satisfied, but **the two conditions carry different weight.** The honest form is not "the error rate is lower" but "**in the same error-rate band (both under 1% by point estimate, indistinguishable)**", and the FOUND side does not separate after cluster correction either.

## Appendix J — governor re-adjudication, 40 items (results)

Re-adjudication of the 40 items in `HANDCHECK_SAMPLE_2026-09-12.json` **with the source opened**. Sample pre-registration = Appendix E.

| Sample | n | Local verdict | Governor verdict | Disagreement |
|---|---|---|---|---|
| O decisive items (NOT-REAL 1 + UNVER 7) | 8 | 1 N · 7 U | 1 N · 7 U | **0** |
| O REAL random | 12 | 12 R | 12 R | **0** |
| F_gen REAL random | 20 | 20 R | 20 R | **0** |

Breakdown: O's UNVER 7 = nonexistent module ×3 · missing tests ×2 · a struct field in another file · a model property — all correctly "outside the file". The NOT-REAL 1 (h01 TOCTOU) uses `dir_fd`-relative operations + `AT_SYMLINK_NOFOLLOW`, so it does not escape as claimed.

**Do not read this 40/40 as "verification" — three things apply at once.** ⓐ **The re-adjudicator is the governor model, not a human** (in the source record's own words, "axis-① level independent re-adjudication — not by human hand"), so this is **a second model's agreement** and does not rule out correlated error between models. ⓑ **40 cannot bound an error rate of the 0.5% class**: for 40 items with no disagreement the 95% upper bound on the disagreement rate is ≈**7%**, so this sample's precision is **one order of magnitude coarser** than the effect it is meant to warrant. ⓒ **The sample composition is one-sided** — F_r2ctrl 0 items · F_typed 0 items · 0 items in the NOT-REAL direction for the FH arms — therefore **the ruler's "rejection" direction is unverified. And this re-adjudication is a post-hoc promotion, not a prior calibration** (§3.3) — it took place at 01:0x the same day, **after** the F_typed·F_gen·O scoring had finished. It must not be read as "the ruler was verified first". This section of the original record writes the seed count as "88/88", but **that number is not the total** (§4 correction: per arm 56/60/62/28, total **206**; 88 = F_gen 60 + O 28). This appendix follows the corrected values.

---

# Appendix K — The record of narrowing

> Quoted wording in this record is translated; the Korean version of this paper carries the original strings.

This appendix is the record of what this study narrowed, and when. Since, as §10.1 says, the **order** of corrections is this paper's contribution, the correction history is not erased but collected here. Each "(Appendix K.n)" in the body points to this list. The narrowing happened at four places: ① cluster correction · within-run contrast · positive controls · the A-6 scoring table · the sealed-consistency label · 2-pass cost (2026-09-12/13) · ② the 3-axis repair from adversarial-review round 2 (2026-09-13) · ③ the remaining R items from the same round plus the contributions section (2026-09-13) · ④ incorporation of the prior-work survey (2026-09-17 — §2 added · §10.4 grounds replaced · Appendices H·I·J attached). "Adversarial review R-n" is the finding number from the 2026-09-13 cross-family adversarial review, round 2 (codex · agy) — R1 possible unrecorded disposal of non-seed claims · R2 prompt byte-identical ≠ design-identical · R3 exchangeability assumption of the permutation test unstated · R4 resolution ≠ power · R5 no multiple-comparison correction applied · R6 the three metrics' prediction failures bundled into one cause · R7 transfer limit of the positive controls · R8 the "family effect" reading in §5 · R9 the claim in §5 that external citations support the direction · R10 the estimator of the wall-clock multiple · R11 absence of an input-ID ↔ output-verdict-ID correspondence · R12 circumstantial inference about sealing. Of these, R1 · R7 · R11 are items that require a new experiment rather than an edit, which is why the body records those designs only as commitments.

## K.1 Title — 2026-09-12/13

The draft title was "Doubles planted-defect detection without paying error rate". "Doubles" does **not stand** after cluster correction (§4.2) and the within-run contrast (§4.1) — no contrast is significant, and run-to-run variance under an identical design exceeds the effect. What stands is "does not subtract" (paired losses 0/24 · 0/24) and the anatomy of the three "free" paths. The title was changed accordingly.

## K.2 §1 General proposition — 2026-09-13

We first wrote §1 as "discarding more candidates raises precision". That general proposition is false (discarding mostly correct candidates lowers precision), so it was changed to "if the discard criterion is correlated with 'likely to be wrong'".

## K.3 §3.2 Number of arms — 2026-09-12/13

We first wrote "five arms", a number that appears in no source. The arms of this study were corrected to four — `N`·`F` are baselines from the prior measurement and were not re-run.

## K.4 §3.3 Order of calibration and measurement — 2026-09-12/13

We first wrote the order as "calibrated first, then measured the arms". The seeds are simultaneous with the arm measurement, and the governor re-adjudication is a post-hoc promotion after scoring ended, so that order claim was corrected.

## K.5 §3.3 ⓒ Defense deleted — 2026-09-13

We first wrote, as ⓒ: “applying the same key to the round-1 output gives F_r2ctrl 20/24, a value that could not arise if the key had been fitted only to reproduce 18 and 14”. That does not hold: a key overfitted to the calibration sample can produce any total on other inputs, and “a different number came out” is not “it agreed with an independent ground truth.” The overfitting refutation is done by ⓓ (whose ground-truth source is independent); ⓒ was not grounds, so it was deleted.

## K.6 §3.4 Four sealed prediction lines — 2026-09-13

We first omitted the “per-tier” row and wrote “one prediction was wrong”. §5.1 had already corrected A-6 with “quoting only one line is selective reporting”, and Appendix G then did the same thing. Corrected to four lines · two wrong.

## K.7 §3.4 Two pre-registration deviations — 2026-09-13

The work record wrote “recorded by name”, yet we did not carry these two deviations over. We also described the “alternation” in ② as a “defect” of the first fixture, but the alternation was not a defect but a sealing requirement — that fixture’s real defects are two: id leakage and submitting mutant and original together (a violation of pre-registration §2).

## K.8 §4 Definitions in the results table and the coverage column — 2026-09-13

We first published three cells of the results table (error rate · non-REAL · coverage) without definitions, and defined coverage as the complement of the abstention rate, a definition that did not match its own table — the three FH values in the table were pass rates of the pipeline’s verification layer, and only the O value was the scorer’s abstention rate. That is, one column was mixing a different instrument per arm. The column was split into two (verification pass rate · adjudication rate), and for the same reason the coverage-axis test we had published (a value comparing the two instruments) was retracted from §4.

## K.9 §4 Seed count — 2026-09-12/13

We first wrote “seeds 88/88”, but that 88 is **F_gen 60 + O 28**, not the whole. Written per arm, with the sum in the §4 table.

## K.10 §4 The O-arm empty-run sentence — 2026-09-13

The sealed Appendix D instructed that “the paper states that O’s values are partly the product of slots with no prose”, and we omitted that sentence. §4 now carries that sentence, the number of empty runs, and the FOUND upper bound.

## K.11 §4 O-arm mitigating circumstantial inference — 2026-09-13

We attached one more sentence here: “a document rewritten after seeing the results has weak reason to leave a refuted prediction in place”. That mitigating inference was retracted (adversarial review R12).

## K.12 §4 Coverage argument — 2026-09-13

We used “coverage is higher for octo” as an argument here, but that comparison itself did not hold (two instruments). It was retracted, and the fact that the competing explanation (the extraction-artifact hypothesis) was not retracted on the coverage axis was restored.

## K.13 §4.1 Byte-identical ≠ design-identical — 2026-09-13

We first used “prompt byte-identical” and “design-identical” as the same statement (adversarial review R2). The precise statement was lowered to *“the stored round-1 prompts are identical”*.

## K.14 §4.2 Number of extreme cells — 2026-09-13

We first wrote “18 of 24 cells” here — a number that comes out of no cut (21 of 32 · 17 of 24). The working results record was wrong in the same way and was corrected together. The argument stands on either value.

## K.15 §4.2 Test-assumptions paragraph — 2026-09-13

This paragraph was missing entirely at first — test statistic · null hypothesis · assignment scheme · threats to exchangeability (adversarial review R3). It was added.

## K.16 §4.2 Multiple comparisons — 2026-09-13

We only stated here that “applying multiple-comparison correction would move it further away” (adversarial review R5). It was actually applied and the Holm·Bonferroni values were put in the table.

## K.17 §4.3 Added — 2026-09-13

§4.3 was added while fixing two errors in the sentence we first wrote (reading the lower bound as power · “5 wins 0 losses”) (adversarial review R4). The history is left in the body as is — because what was wrong is the content of the section.

## K.18 §4.4 · §4.5 Added — 2026-09-13

§4.4 was added on 2026-09-13 to answer adversarial review R1 (“it may have dropped other claims while still catching the seeds”), and §4.5 to survey the prior literature on that gap. The retraction of that §4.4 paragraph (“the big hole”) is from

the same day and is left in the body as is — that retraction is the section's argument.

## K.19 §4.5 Motivation sentence — 2026-09-13 (same day)

We first wrote here: "these 4 values eliminate our '113 unrecorded disappearances' — what we lacked was a place to give a rejection a type", but that rests on a premise retracted in §4.4. The exact gap is the absence of claim-level types. The value of the survey does not shrink, but its reason changes.

## K.20 §7 #5 Calibration-table reference — 2026-09-13 · attached 2026-09-17

We first cited this table (from its working results record) as "Appendix 3 §10", and that number does not exist — in the submission it is actually attached as Appendix H.

## K.21 §7 #8 Windows and the selection rule — 2026-09-13

We first reduced this item to "time-of-day effect unmeasured" only. The number of windows · dead runs · the per-(case,rep) "last successful run" selection rule were stated explicitly.

## K.22 §7 #10 "Uncorrected" conflict — 2026-09-13

We first wrote "uncorrected" on this item, conflicting with its own §4.2 — a line carried over from a discarded working repair list. Corrected.

## K.23 §8 Number of defects — 2026-09-13

We first listed only three defects (①②③). ④ the preflight looked at only one side of the roster · ⑤ the error channel was in fact the transcript were added, making five.

## K.24 §8 ③ The two dead units — 2026-09-13

We first wrote "the two dead ones are the units with the most", which is true of only one.

## K.25 §8 ⑥ 2-pass cost — 2026-09-13

Our first conclusion on §8 ⑥ (the wall-clock multiple exceeds the pre-registered prediction) and its "cause" explanation (the counterpart's list is attached, enlarging the prompt) were retracted (adversarial review R10). The two layered instrument errors (E1 · E2) and the re-measurement history are left in the body as is — that history is itself an operational harvest. We also placed this item beside the three misses of §5.1 as "a fourth miss in a different direction"; since the prediction was right and the measurement was wrong, that placement was retracted.

## K.26 §8 Budget figures and framing — 2026-09-13

We first bundled the three budget figures into one parenthesis and laid a discarded framing on top. Two framings retracted: ① "**daily** bucket" — that family's reset is **weekly** (the work record explicitly corrected this the same day, and the draft reverted it). ② "**2 or more windows**" — the correct conclusion is "even one weekly reset may not suffice".

## K.27 §9 Narrowing of the conclusion — 2026-09-13

The three narrowings §9 made against what we first wrote — "round-2 does not subtract" as an observation rather than a guarantee · retraction of the assertion "there is no value in increasing reps" · separating the evidence grades of the three "free" paths — are the argument of the conclusion, so **the narrowed claims themselves remain in §9.** The record of the narrowing — what we first wrote and how it was corrected — is carried here: §9 writes "what stands", this appendix writes "how it narrowed". Stating a correction against the deleted sentence, inside the conclusion, asks a reader who does not have that draft to read a diff instead of a position.

## K.28 §10.4 Grounds replaced — 2026-09-17

The retraction of the §10.4 grounds we first wrote ("cost structure → cannot be promoted") is the argument that motivates §10.4.1, so it stays in the body. The same reasoning was also in abstract ⓓ and was replaced the same day. The grade table in the §10 introduction and §2 were also added the same day — how the prior-work survey narrowed the contributions is recorded in the §10 introduction.

## K.29 Appendix sections — delegation of numeric authority to the work record — 2026-09-17 (accumulated)

The submission cannot say "on disagreement, the results record is canonical" — that file is not part of the submission, and delegating numeric authority to a document the reader cannot check is an unverifiable claim. We first wrote it that way. Instead, each appendix has its own sealed file, and the body's figures carry definitions and recomputations together so that they close within the body.

But that sentence alone did not close it — because the correction landed "beside" the original sentence rather than "replacing" it. A lower line in the same section still delegated numeric authority to that file. So the dependencies were **counted exhaustively and split into two kinds** (exhaustive grep, control = 0 hits for a nonexistent token).

| Kind of dependency | Location | Disposition at submission |
|---|---|---|
| ⓐ **Correction history** — "what we wrote and how we fixed it" | §1 pre-registration deviations · §4.1 18→17 correction · §5 arm provenance · §5.1 post-hoc-interpretation marking · | **Kept as is** (in the body or in this Appendix K). The reader can verify the paper's figures without that file, and these |

| Kind of dependency | Location | Disposition at submission |
|---|---|---|
| | §5.1 three-way split not restored · §6 adjudication-rule correction · §7 #7 pairing not run to completion · §8 window-framing retraction | citations point to **history**, not to numeric authority |
| ⓑ **Delegation of numeric authority** — without that file the value cannot be checked | **Only three**: ① the header's "numeric basis = RESULT §12~§21" ② §7 #5's "the bidirectional calibration table is in results record §9" ③ the three items of the "Appendix pointers" section (bidirectional calibration table · O adjudication application · governor re-adjudication 40 cases) | **Must be closed before submission.** ① is a working-draft notation and is deleted in the submission. ② — the body **already carries inline** the values it needs (n=6 · all gemini·security), so the pointer is demoted to history. ③ is **actually attached as Appendices H·I·J** (2026-09-17) |

Thus the three delegations were closed: ⓑ① deleted from the submission, ⓑ② demoted to history (Appendix H attached), ⓑ③ **executed by attachment**. The sentence originally in this place, "at submission it must be one of the two", was a **deferral**, not a disposition, and we wrote that deferral as if it were a disposition. What remains now is one cell: the **absence of a sealed hash** for the three results tables. That is a property, not a gap — results tables are not pre-registrations, so "fixed before seeing results" does not apply to them.

The figures of the three outputs (bidirectional calibration table · O adjudication application · governor re-adjudication 40 cases) existed only in the working results records that are not part of the submission (the source records of Appendices H·I·J respectively). The options were ⓐ **actually attach** as appendices · ⓑ remove the body claims that depend on those three, and ⓐ **was executed** — for Appendix J, §3.3·§3.5 carry the numbers inline (composition of the 40 cases · 40/40 · the 95% upper bound ≈7% self-limit argument), so ⓑ was impossible from the outset. Appendices H·I were both open, and attachment was chosen.

The names were also split — Appendix E already carries the label "governor re-adjudication 40 cases", but what is attached there is the **pre-registration** of that re-adjudication (`PREREG_ADDENDUM5`), and what was missing was that re-adjudication's **results table**. Calling them by the same name leaves "already attached" and "missing" indistinguishable. So Appendices H·I·J each point to a **results table**: Appendix H = bidirectional calibration table · Appendix I = O adjudication application · Appendix J = governor re-adjudication **results** 40/40.

## K.30 Two values not carried over when attaching Appendices H·I·J — 2026-09-17

While carrying them over, two things were deliberately not carried — because if an appendix holds a number older than the body, **that place becomes the source of the next correction**: ⓐ the coverage row in Appendix I's source record and its `p<0.001` (retracted by the body · a comparison of two instruments) ⓑ the "seeds 88/88" in Appendix J's source record (the sum is 206 and 88 is F_gen 60 + O 28). The source records still hold both, and **since those files are not part of the submission they are not corrected.** The next revision must not revert them as "missing".

## K.31 Appendix instrument list — 2026-09-13

The instrument list we first published was not "exhaustive", and the producers of the numbers cited by the body and abstract were missing — of the directory's full set of instruments (lanes excluded), only 11 were listed. The eight that produce the cited numbers were added. The adversarial review counted this omission as "5 kinds", which was an undercount — 8 kinds of numeric producers alone, 22 kinds in total.

## K.32 Correction blocks left in the body — —

The following corrections·retractions are the argument of their sections themselves, so they are not moved to this appendix and are left in full in the body: §4.3 (resolution ≠ power, "5 wins 0 losses") · §4.4 (retraction of the "big hole" paragraph — "it flipped when we opened another channel") · all of §5 and §5.1 (retraction of "generation-time decorrelation" · correction of "separated" · restoration of the post-hoc-interpretation marking · three-way split not restored · separating the provenance of the sealed misses · correction of the logical error "the three misses fold into one") · §6 (overlap/unique decomposition not reported, unrecoverable · single observation downgraded · retraction of the claim that external citations support the direction) · §8 ⑥ (wall-clock re-measurement) · §10.4 (retraction of the cost-structure grounds) · §10.4.1 (the "what is not claimed" table).

## K.33 §7 #16 multiple-comparison wording — 2026-09-18

§4.2 has actually applied Holm·Bonferroni since the Appendix K.16 revision, but §7 #16 still carried the sentence we first wrote, "no multiple-comparison correction was performed" — the English translation arm flagged the tension with §4.2. Narrowed to "not applied paper-wide — §4.2 applies it only within the family of 4". No number changed.

---

## K.34 §10.3 heading · the unit of the instrument count · where the paired arms belong — 2026-09-18

A zero-context reviewer read-through of the English build named three places where the body contradicted itself; these are source-side problems, so both language versions were fixed together. No number changed. ⓐ Only the §10.3 heading said "decodability" — the body, the abstract and the §10 grade table all say "decomposable", and that word occurs exactly once in the paper. Unified to "Decomposability". ⓑ "33 instruments" and K.31's "22 kinds in total" stood side by side with nothing reconciling them — 33 counts files, 22 counts kinds. The unit is now stated in both places. ⓒ `F_pair_blind` / `F_pair_open` appeared in the §4.4 table with no definition after §3.2 declares four arms — a clause now says they are the paired arms of Appendix F and did not run to completion (§7 #7). The five first-person passages (§3.4 · §4.3 · §4.4 · §6) were KEPT. The Korean "I/my" carries the personal attribution of an instrument error, and that is §10.1's argument; rewriting it as "we" weakens it.

## K.35 §10.6 added — the name of the practice this method belongs to — 2026-09-18

The paper had never named the practice it belongs to. As a result the reader had to infer that §10.1 (fixing the ruler first and letting it kill our own hypothesis) and §8 (the three paths by which a metric improves for free) are **two faces of one discipline**. §10.6 writes that name down: *our way of AI governance engineering*. **It claims no novelty, and that is half of what the section is for.** Both the term and the field have precedents: [ZWF26] (a policy-enforcement engine inside a harness) · [OC26] (a governance gate opened and closed on signal thresholds) · [BSN26] (accounting for residual violations by the enforcement stack's own error rate) · [PN26] (norms translated into executable pipelines) · [MF22] (the organizational-layer study). All four new entries were **confirmed by reading the arXiv abstract directly** — title, authors, v1 date, subject class and a verbatim span. No entry was taken second-hand from a search-result summary. [ZWF26] reuses a key already present in the references. A limit is carried with it: the failure of step-level compliance to compose, named by [KGG26], is not addressed by this paper. No figure changed (the only numbers §10.6 introduces are the cited prior work's own: 78.6%, 64.9% and 43; nothing of this paper's own figures moved).

## K.36 Title replaced — the title was a sentence K.27 had retracted — 2026-09-18

**Finding**: the title read "A Second Pass Does Not Subtract", and **K.27 (2026-09-13) retracted exactly that sentence** — it judged *"round-2 does not subtract"* to be *turning an observation into a guarantee*, and §9 states that what stands is only *"in this sample we did not observe a loss of planted-defect detection"*. The correction landed in §9 and in K.27 and **never reached line 1** — the retracted claim sat for five days in the one line every reader sees first. The abstract was already calibrated (contrast by contrast, p in two columns), so the title alone disagreed with both the body and the abstract.

**Resolution**: rather than repairing the modality (past tense, "did not subtract"), the clause was **removed**. The new title leads with §10.1, which is this study's methodological contribution — the ruler was fixed first, and it killed the authors' own hypothesis. In a paper whose primary result is null, that presents the null as evidence of the method rather than as a weakness.

**Absorbed into the first clause instead of adding a subtitle**: the cost of dropping "planted defects" from the title was measured first — the term appears 11 times in the English body and the abstract's first paragraph carries it, so the indexing loss is small. The loss falls on a reader scanning a listing, so instead of a new subtitle the phrase was absorbed as "our own hypothesis on planted-defect detection". Title length 118 → 132 characters (arXiv imposes no title-length limit; the 1,920-character limit applies to the abstract FIELD).

The title in the submission metadata was brought into line with it.

## K.37 Title replaced again, and two claims the abstract made that the body had already withdrawn — 2026-09-19

**Finding 1 — half-retraction, coverage axis.** §4 states that *"the between-arm comparison on the coverage axis does not hold, and that axis's test `p<0.001` is retracted (it was a value comparing two instruments)"*, and §4 separates the two instruments by name: 92.6% is an **adjudication rate**, while 67.2 / 68.5 / 65.6% are *"the pass rate of the pipeline's verification layer."* The abstract and §10.3 nevertheless carried the comparison as fact — §10.3 stating the direction outright (*"octo, at 92.6%, is higher than we are (68.5%)"*). Both were brought into line with §4. 🟥 **No counter-claim was put in its place**: the figures on a single instrument (F_gen 99.4% vs O 92.6%) are *not* asserted, because that is a different claim and this appendix is not the place to make one. The numbers themselves stay where §4 defines them; only the instrument-free *comparison* is gone. §10.3's argument does not rest on it — the section is titled *"Is Not a Number but Decomposability"*, and what is one-sided there is the record, not a ranking.

**Finding 2 — half-retraction, the 95% value.** §9 states *"we do not cite it as an upper bound on the actual error rate"* until clustering is reflected. The abstract cited it as an upper bound **and reasoned from it** (*"since the upper bound exceeds octo's point estimate of 1.1%…"*). That inference is withdrawn. 🟥 The value is **kept, not deleted**, in the §4 cell with §9's qualifier attached — §9 asked that it not be *cited as an upper bound*, not that it be removed, and §3.5 explicitly permits stating the point estimate together with its interval. Deleting it in one place and leaving it in another is how a reader later fails to tell which of the two is stale.

**Finding 3 — the title claimed a cause the body assigns elsewhere.** The 2026-09-19 replacement first read *"…and **the Positive Control** Killed Our Own Planted-Defect Result."* The body does not support that attribution: §3.4's pre-registered positive controls **established the ruler's discrimination** (fabricated claims rejected 12/12), and what removed the result was **cluster correction** — §4.2, *"After cluster correction, FOUND is p=0.109, 'not separated'."* The title now names that. 🟥 The first replacement was a compression of §10.1's chain (*"when we measured our own *main effect with that ruler, nothing remained"*), and the compression moved the cause by one link. Recorded because it is the same failure mode this appendix's K.36 entry was written about — **a title asserting more, or other, than the body.**

**Also corrected here**: K.36 states *"Title length 118 → 132 characters"* and that the phrase was absorbed as *"our own hypothesis on planted-defect detection."* Both describe the 09-18 title and are now historical: the current title is **104 characters** and contains neither phrase. K.36 is left unedited — a changelog rewritten to match the present is no longer a record — and this entry supersedes its final line: **the submission metadata must be brought into line with the title above, not with K.36's.**

No figure changed.